\newif\ifarxiv
 \arxivtrue 

\ifarxiv

\documentclass{article}

\newcommand{\rapattack}{\textsc{RAP-Rank}}
\newcommand{\rap}{\textsc{RAP}}

\newcommand{\matchrate}{\textsc{Match-Rate}}

\newcommand{\indicator}[1]{\mathbbm{1}\left\{#1\right\}}

\newcommand{\cX}{\mathcal{X}}

\usepackage{fullpage}
\usepackage{bbm}
\usepackage{enumitem}
\usepackage{float}
\usepackage{color-edits}
\usepackage{algorithm}
\usepackage{algpseudocode}
\usepackage{authblk}
\usepackage{amsfonts}
\usepackage{amsmath}
\usepackage{hyperref}
\usepackage{booktabs}
\usepackage{graphicx}
\usepackage{multirow}
\newtheorem{theorem}{Theorem}
\newtheorem{lemma}{Lemma}
\newtheorem{definition}{Definition}

\title{Confidence-Ranked Reconstruction of Census Microdata\\ from Published Statistics}

\author[a]{Travis Dick}
\author[b]{Cynthia Dwork} 
\author[c]{Michael Kearns}
\author[d]{Terrance Liu}
\author[c]{Aaron Roth}
\author[e]{Giuseppe Vietri}
\author[d]{Zhiwei Steven Wu}

\affil[a]{University of Pennsylvania (Now at Google)}
\affil[b]{Harvard University}
\affil[c]{University of Pennsylvania}
\affil[d]{Carnegie Mellon University}
\affil[e]{University of Minnesota}

\date{}

\else

\documentclass[9pt,twocolumn,twoside,lineno]{pnas-new}

\usepackage{xr-hyper}
\newcommand{\rapattack}{\textsc{RAP-Rank}}
\newcommand{\rap}{\textsc{RAP}}

\newcommand{\matchrate}{\textsc{Match-Rate}}

\newcommand{\indicator}[1]{\mathbbm{1}\left\{#1\right\}}

\newcommand{\cX}{\mathcal{X}}

\makeatletter

\newcommand*{\addFileDependency}[1]{
\typeout{(#1)}
\@addtofilelist{#1}
\IfFileExists{#1}{}{\typeout{No file #1.}}
}\makeatother

\newcommand*{\myexternaldocument}[1]{%
\externaldocument{#1}%
\addFileDependency{#1.tex}%
\addFileDependency{#1.aux}%
}
\myexternaldocument{SI_standalone}

\usepackage{xcolor}
\usepackage{bbm}
\usepackage{enumitem}
\usepackage{hyperref}
\usepackage{float}
\usepackage{ulem}
\usepackage{color-edits}

\usepackage{multirow}
\usepackage{algorithm}
\usepackage{algpseudocode}

\newtheorem{definition}{Definition}

\addauthor{sw}{blue}

\templatetype{pnasresearcharticle} 

\title{Confidence-Ranked Reconstruction of Census Microdata from Published Statistics}

\author[a]{Travis Dick}
\author[b]{Cynthia Dwork} 
\author[c,1]{Michael Kearns}
\author[d]{Terrance Liu}
\author[c]{Aaron Roth}
\author[e]{Giuseppe Vietri}
\author[d]{Zhiwei Steven Wu}

\affil[a]{University of Pennsylvania (now at Google)}
\affil[b]{Harvard University}
\affil[c]{University of Pennsylvania}
\affil[d]{Carnegie Mellon University}
\affil[e]{University of Minnesota}

\leadauthor{Dick} 

\significancestatement{We show how to launch a reconstruction attack on US Decennial Census data based only on publicly released statistics. The attack can recover individual microdata --- i.e. the responses of individual Census survey respondents. Although our attack cannot reconstruct all rows of the private data, it can produce a confidence-based ranking of rows, such that rows that appear earlier in the ranking are more likely to appear in the private data. Thus the attacker can reconstruct a fraction of the rows with confidence. We compare our attack to a hierarchy of increasingly strong baselines, and show that we can outperform all of them. Our results point to the necessity of employing privacy enhancing technologies when releasing statistics of privacy-sensitive datasets.}

\authorcontributions{All authors contributed to all aspects of the research and writing. Order of authors is alphabetical.}
\authordeclaration{This submission deals with reconstruction of census data, and specifically addresses questions related to the U.S. Census Bureau. Reviewer Kunal Talwar is a member of the Census Scientific Advisory Committee (\url{https://www.census.gov/about/cac/sac.html}), which is a committee consisting of subject matter experts that advise the Bureau on technical matters. Talwar does not receive any compensation or honoraria for this work.}
\correspondingauthor{\textsuperscript{1}To whom correspondence should be addressed. E-mail: mkearns@cis.upenn.edu}

\keywords{Dataset Reconstruction $|$ United States Census $|$ Privacy}

\dates{This manuscript was compiled on \today}
\doi{\url{www.pnas.org/cgi/doi/10.1073/pnas.XXXXXXXXXX}}

\fi 

\begin{document}
\maketitle
\begin{abstract}
A reconstruction attack on a private dataset $D$ takes as input some publicly accessible information about the dataset and produces a list of candidate elements of~$D$.
We introduce a new class of data reconstruction attacks based on randomized methods for non-convex optimization.
We empirically demonstrate that our attacks can not only reconstruct full rows of $D$ from
aggregate query statistics $Q(D)\in \mathbb{R}^m$, but can do so in a way that reliably ranks reconstructed rows by their odds of appearing
in the private data, providing a signature that could be used for prioritizing reconstructed rows
for further actions such as identify theft or hate crime. 
We also design a sequence of {\em baselines} for evaluating reconstruction attacks.
Our attacks significantly outperform those
that are based only on access to a public {\em distribution} or population from which the private dataset $D$ was
sampled, demonstrating that they are exploiting information in the
aggregate statistics $Q(D)$, and not simply the
overall structure of the distribution. In other words, the queries $Q(D)$ are permitting reconstruction of elements of {\em this} dataset, not the distribution from which $D$ was drawn. These findings are established both on 2010 U.S. decennial Census data
and queries and Census-derived American Community Survey datasets. Taken together, our methods and experiments illustrate
the risks in releasing numerically precise aggregate statistics of a large dataset, and provide
further motivation for the careful application of provably private techniques such as differential privacy.
\end{abstract}

\ifarxiv
\else

\thispagestyle{firststyle}
\ifthenelse{\boolean{shortarticle}}{\ifthenelse{\boolean{singlecolumn}}{\abscontentformatted}{\abscontent}}{}
\fi


The goals of data analysis and those of responsible data stewardship are often in tension:
we wish to extract and share
useful information about important datasets, 
but also must maintain the privacy of the individuals whose information comprises the datasets. 
This is exactly the problem faced by the U.S. Census Bureau, which on the one hand has a legal mandate to protect the privacy of its respondents,\footnote{Title 13, Section 9, of the US Code: says "(a)  Neither the Secretary, nor any other officer or employee of the Department of Commerce or bureau or agency thereof, or local government census liaison, may, except as provided in section 8 or 16 or chapter 10 of this title [13 USCS § 8 or 16 or §§ 401 et seq.] or section 210 of the Departments of Commerce, Justice, and State, the Judiciary, and Related Agencies Appropriations Act, 1998 [13 USCS § 141 note] or section 2(f) of the Census of Agriculture Act of 1997 [7 USCS § 2204g(f)]— 
(1)  use the information furnished under the provisions of this title for any purpose other than the statistical purposes for which it is supplied; or
(2)  make any publication whereby the data furnished by any particular establishment or individual under this title can be identified; or
(3)  permit anyone other than the sworn officers and employees of the Department or bureau or agency thereof to examine the individual reports."
} but on the other hand is a major governmental statistical agency that released over 150 billion tabulations from the data collected as part of the 2010 decennial Census \cite{abowd}. 
 What is the risk of releasing large numbers of aggregate statistics from this and similarly 
sensitive datasets?

 In this work, we empirically demonstrate that without taking explicit and rigorous steps to ensure individual privacy, the release of simple aggregate statistics of large datasets is highly vulnerable to specific and computationally feasible attacks that can reliably reconstruct complete rows of the private dataset. These findings stand even when measured against a hierarchy of baseline metrics that correspond to ``reconstructions'' that result from increasingly fine-grained knowledge of the distribution from which the private data was drawn. 
 
 {\bf Confidence.}
 An important and damaging aspect of the attacks we propose is their ability to give confident predictions about which rows have been correctly reconstructed when the reconstructions are not perfect (as is typically the case). Our methods output a {\it ranking} over candidate reconstructed rows that is highly correlated with their
presence in the private data, with rows that appear early in the ranking having high odds of appearing in the true dataset. Such a ranking could be used by an adversary to prioritize subsequent exploitation of private data --- for example, for identity theft, or to locate individuals of certain backgrounds (or even in the intersection of certain background, age range, and sex categories).\footnote{Census data were used to identify Japanese Americans to send to internment camps during World War II \cite{washingtonpost, SeltzerAnderson}. In 2002, the Department of Homeland Security used re-tabulations of census data to track Arab-Americans \cite{nytimes}.}

Our methods are based on casting the reconstruction problem as an instance of large-scale, non-convex optimization, along with a subsequent step to convert non-continuous
(e.g. categorical) features back to their original schema. The algorithms are closely related to recent methods for generating synthetic 
versions of sensitive datasets \cite{ aydore2021differentially,liu2021iterative,vietri2022private} while enforcing Differential Privacy (DP)~\cite{dwork2006calibrating, DBLP:conf/icalp/Dwork06, dwork2016calibrating,dwork2014algorithmic}. Crucially, these techniques are randomized, which allows us to repeat the reconstruction multiple times and get different results. Our confidence rankings are obtained by computing how frequently rows appear in multiple reconstructions; this is the primary methodological novelty of our approach.\footnote{Prior deployed data reconstruction attacks are generally deterministic and based on linear or integer programming. Theory for such attacks dates back to \cite{dinur2003revealing} (see \cite{kasiviswanathan2010price} for theory for attacks on release of contingency table statistics, which closely match the kinds of statistics released by Census that we attack here). See \cite{dwork2017exposed} for a survey of privacy attacks, and \cite{cohen2020linear} for a description of a practical linear programming-based reconstruction attack. Integer programming techniques could also be used in our case together with a randomized objective function to provide the needed stochasticity. We investigated this as well, but found that the continuous optimization approach was more performant --- especially on higher dimensional data like data from the American Community Survey.} 

We establish the striking and consistent empirical phenomenon that rows that appear more frequently across repeated optimizations are also more likely to be rows of the private dataset --- thus allowing an attacker to build confidence in the authenticity of rows that are repeatedly reconstructed, and to prioritize individuals for other forms of attack. We present a simple mathematical justification in a Bayesian framework that provides intuition for this phenomenon. We remark that reconstruction (recovering the rows of the private dataset from published statistics) is a different problem from re-identification (re-attaching names to the reconstructed rows) --- but prior work \cite{rocher2019estimating} shows that re-identification risk can be confidently estimated given reconstructed rows, and is generally high for data with even a moderate number of recorded features.


{\bf Baselines.} The Census bureau itself conducted a reconstruction attack on 2010 Census data, and reported top-level statistics on what fraction of rows in the reconstruction matched rows in the original data --- 46\% \cite{abowdreport}. This has been dismissed as benign because the top-level statistics were not compared to 
baseline ``attacks'' that could be
performed from publicly available data, and which might have achieved similar top-level match rates \cite{rugglesreport,ruggles2022role}. 
While the criticism is simply incorrect in the case of small Census blocks~\cite{privacyexperts}, it does raise the important question of choosing appropriate baselines, and what we should measure, beyond top-level reconstruction rates, to indicate when we should view reconstruction attacks as worrisome. For example, even if we cannot reconstruct all or even most rows in a dataset, can we have confidence that some of our particular reconstructed rows are correct? If so, this is reason for concern even when top-level reconstruction rates are well below 100\%.

Thus a key aspect of our work is explicit comparison to distributional baselines that correspond to increasingly precise knowledge of the distribution from which the sensitive dataset was sampled (but not which rows specifically were sampled). We introduce a hierarchy of increasingly challenging baselines
that correspond to the ability to sample datapoints from the true distributions of individuals at successively finer geographic census units --- namely, sampling at the national, state, county,
tract and block levels. Our attacks are based on an optimization subroutine that is initialized with some seed dataset, which is then iteratively improved. Even when we start with a uniformly random initialization, our attacks consistently outperform all but the most granular baseline in terms of the number of private rows
perfectly reconstructed. 
If we instead initialize our attack at the baseline distribution to which we compare --- in accordance with
the assumption that the distribution is already publicly known --- then we consistently outperform even the most granular baselines, according to this top-line metric.
In other words, the effectiveness of our attack is significantly amplified above any baseline distribution by access to that baseline, establishing that 
we are significantly  
exploiting additional information revealed by the aggregate private dataset statistics.


Our primary experiments are performed on synthetic U.S. Census microdata released by the Census Bureau during the development of the 2020 Census Disclosure Avoidance System (DAS), together with the queries corresponding to the actual statistics published by the U.S. Census Bureau at various levels of geographic granularity. In particular, we use queries corresponding to the same tables that the Census Bureau used in their internal reconstruction attack of the 2010 Census data~\cite{JASONPrivacyMethods2022}. 
To demonstrate the generality of our findings, we also run our reconstruction attacks on large, significantly higher dimensional datasets derived from the Census Bureau's American Community Survey, as processed via the Folktables package \cite{ding2021retiring}. In addition to demonstrating that the attacks scale to larger (and potentially more sensitive datasets), this allows us to deliberately control the queries released and study phenomena such as the effectiveness of the attack as a function of the types
and numbers of queries. 

{\bf Consequences.} This work demonstrates that the large-scale, non-convex optimization techniques we employ
can be viewed as both a ``poison'' and a ``cure''. Applied carefully to {\it noisy} statistics that ensure DP, they provide a powerful tool for sharing useful data statistics while ensuring citizen privacy \cite{ aydore2021differentially,liu2021iterative,vietri2022private} . But applied to numerically precise statistics --- even those that may provide only aggregate information about large populations --- they exfiltrate entire rows of sensitive data with confidence.

In the context of the Census, our findings yield sober warnings on the privacy risks of releasing
aggregate queries without explicit safeguards such as Differential Privacy. Our attacks use numeric values of released statistics of the underlying Census microdata, and are agnostic to how these are encoded. So, for example, the vulnerability of the data is in no way reduced if instead of releasing tabular statistics, ``synthetic microdata'' are released (as has recently been proposed for the American Community Survey \cite{fullysynthetic}) that are consistent with the underlying tabular statistics; our attack can be launched against such synthetic microdata by simply computing the underlying statistics from the microdata. The only defenses against our attack are to introduce imprecision in the underlying statistics themselves, as techniques like differential privacy do. 

We note that all of our experiments were
performed on standard, commercially available consumer compute hardware,\footnote{Experiments were run using a desktop computer with an Intel Core i5-4690K processor, 16GB of RAM, and an NVIDIA GeForce GTX 1080 Ti graphics card.}
highlighting the ease and potential prevalence of such attacks. All code and data for replicating our experiments can be found here: \href{https://github.com/terranceliu/rap-rank-reconstruction}{https://github.com/terranceliu/rap-rank-reconstruction}.


\ifarxiv
\section{Preliminaries}
\else
\section{Materials and Methods/Preliminaries}
\fi

In this section, we describe our algorithm, metrics and the baselines we use for comparison.

\subsection{Reconstruction Attack}\label{sec:reattack}

A {\em dataset} is a multiset of records from a discrete domain  $\cX$.  Each item in the multiset is called a {\em row}. 
We use $D\in \cX^\star$ to denote a private dataset that is the target of a reconstruction attack. A reconstruction attack takes as input aggregate statistics computed from dataset $D$ (and in the case of the attacks we present, a possibly uniformly random seed dataset),
and outputs a set of candidate rows, ranked according to the confidence of appearing in $D$. This \textit{confidence-ordered} set of rows is denoted $R=\{R_i\in \cX\}$, where the index~$i$ in $R$ determines the confidence ranking\footnote{Thus $R_1$ corresponds to the row that we are most confident in, $R_2$ the row that we have the next most confidence in, and so on.}.  Our rankings will be obtained from attacks that produce a multiset $X$ of rows. Elements appearing in~$X$ are then ordered according their frequency in the multiset.  We let $R(X)$ denote the resulting ordered set (not multiset) of rows.

To measure the performance of a reconstruction attack, we introduce the following metric that measures its accuracy at different confidence thresholds. For any private target dataset  $D$, the \emph{top-$k$ match-rate} of the confidence set $R$ is the fraction of rows ranked from $1$ to $k$ by $R$ that actually appear in $D$.
\begin{align}\label{eq:matchrate}
    \matchrate_{D,k}(R) \leftarrow \frac{1}{k}\sum_{i=1}^k \indicator{R_i\in D}
\end{align}
We can plot  $\matchrate_{D,k}(R)$ as a function of $k$ which traces out a curve --- in general, if our confidence set has its intended semantics (that higher ranked rows are more likely to appear in $D$), then the curve should be monotonically decreasing in $k$. For a given level $k$, a higher match rate corresponds to higher confidence that rows ranked within the top $k$ are correct reconstructions; at a given match rate, higher values of $k$ correspond to the ability to confidently reconstruct more rows.

Since $R$ is an ordering of unique points, a match rate of 1 even at the highest value of $k$ does not mean that the multiplicity of the points in the original dataset have been reconstructed, but only their identities. In isolation, it is also not clear whether any particular value of $\matchrate_{D,k}(R)$ is cause for concern or not --- whether a given match rate is indicative of a privacy violation or not depends on (among other things) the entropy of the underlying data distribution and on the size of the dataset being reconstructed. As we discuss in Section \ref{sec:baseline}, our view is that $\matchrate_{D,k}(R)$ should be evaluated relative to explicit \emph{baseline} match rates computed on the same dataset being attacked, using statistical information about similar datasets. This comparison shows how much the match rate increases through explicit use of statistics computed from $D$, compared to a (statistically) informed baseline that has no explicit access to $D$, independently of what the size of $D$ is.

\subsection{Reconstruction From Aggregate Statistics}\label{sec:RAP}

We design a reconstruction attack that starts from a collection of aggregate statistics computed from the private dataset. A statistical query is a function that counts the fraction of rows in a dataset that satisfy a given property. We give a formal definition here:

\begin{definition}[Statistical Queries \cite{SQ}] Given a function $\phi \colon \cX\rightarrow \{0,1\}$, a statistical query (also known as a linear query or counting query) $q_\phi$ is defined as  
$q_\phi(D) = \frac{1}{|D|} \sum_{x\in D} \phi(x)$, for any dataset $D$. 
\end{definition}

We use $Q$ to denote a set of $m$ statistical queries and $Q(D) \in \mathbb{R}^m$ to denote the vector of statistics on the dataset $D$. The objective of an attack on $D$ is to reconstruct rows of $D$ given $Q$ and $Q(D)$.

We propose a new reconstruction attack mechanism \rapattack~that learns rows of the unknown dataset $D$ from statistics $Q(D)$. 
\rapattack{} leverages the recent optimization heuristic Relaxed Adaptive Projection\footnote{In fact, what we call RAP in this paper is the simpler algorithm simply called RP in \cite{aydore2021differentially} --- for ``Relaxed Projection''. We do not use the adaptive step.} (\rap) \cite{aydore2021differentially} for synthetic data generation. {\rap} is a randomized algorithm that takes as input a collection of $m$ statistical queries $Q$ and answers $Q(D)\in \mathbb{R}^m$ (derived from some dataset $D$), 
and outputs a dataset $D'$ by attempting to solve the following optimization objective: 
\begin{align}\label{eq:rapobj}
   \text{arg}\min_{D'}\| Q({D'}) - Q(D)\|_2
\end{align}
using a randomized continuous optimization heuristic. 
\rap{} is initialized with parameter $\theta$, discussed below.  Roughly speaking, $\theta$ is an initial guess for what $D'$ should look like, and can be used to capture  additional distributional information available to the attacker.  In our work, this will either be a uniformly randomly generated dataset of a given schema (corresponding to no additional information) or a dataset believed to have been drawn from the same ``distribution'' as $D$; more on this below.
The notation $\uplus$ is used to indicate union with multiplicities.  For example, if $x$ appears 2 times in $D'_1$ and 1 time in $D'_2$, then it appears 3 times in $D'_1 \uplus D'_2$. 

Our method, \rapattack{}, described in Algorithm~\ref{alg:rapattack},
consists of running \rap{} for $K$ times to produce datasets $D'_1, \ldots, D'_K$ and outputting the confidence set $R(\biguplus_{k=1}^{K} D'_k)$.

\begin{algorithm}[!tbh]
\begin{algorithmic}
\State \textbf{Input:} A set of queries $Q$ and their evaluations $Q(D)$ on some private dataset $D$.  
\State \textbf{Parameters:} number of runs $K$ 
\For{$k = 1, \ldots, K$}
    \State Initialize \rap 's parameter $\theta_k$ (either uniformly or to a baseline dataset).
    \State {Output  $D_k'\sim \rap(Q, Q(D), \theta_k)$}
\EndFor
\State Let $D^* = \biguplus_{k=1}^{K} D'_k$ 
\State \textbf{Output:} Confidence set $R(D^*)$ 
\end{algorithmic}
\caption{Overview of \rapattack}
\label{alg:rapattack}
\end{algorithm}

For the purposes of this paper, it is enough to understand that \rap{} is a randomized algorithm that attempts to find a dataset $D'$ which solves the optimization problem in Equation [\ref{eq:rapobj}]. But briefly, it works as follows: 
\begin{enumerate}
\item Datasets $D'$ are defined over a discrete domain, but \rap{} extends this to a larger continuous domain, and extends the queries $Q$ to be defined and differentiable over this larger continuous domain.
\item Starting from its initialization $\theta$, \rap{} then optimizes for a ``relaxed'' dataset $\tilde D'$ in this continuous domain by using stochastic gradient descent on the (now differentiable) objective in Equation [\ref{eq:rapobj}].
\item Finally, \rap{} randomly ``rounds'' $\tilde D'$ back to a discrete dataset $D'$ over the original domain.
\end{enumerate}
So there are potentially three sources of randomness in \rap{}: A (potentially) random initialization point $\theta$, the randomness of stochastic gradient descent, and the final randomized rounding. When we initialize $\theta$ uniformly randomly (corresponding to no prior information) our attack makes use of all three of these sources of randomness. When we initialize $\theta$ to a baseline dataset, we only make use of the 2nd two sources of randomness. We use baseline datasets that result from sampling Census data at various geographic resolutions. 


Although the performance of \rapattack~as measured by \matchrate~is an empirical finding, \rapattack~is a theoretically motivated heuristic. In particular, if we imagine that when \rapattack~is initialized at a sample $\theta$ from a prior distribution on datasets, it samples a dataset $D'_k$ from the posterior distribution on datasets given the statistics $Q(D)$, then the ranking it constructs would be the correct ranking of points by their (posterior) likelihood of appearing in the true dataset $D$. We briefly elaborate on this theoretical intuition in the next section.

\subsection{Some Theoretical Intuition}\label{sec:bayes}

There is a simple Bayesian argument that provides some theoretical intuition for our resampling method for confidently reconstructing rows of the private dataset $D$. Imagine there is some prior distribution $P$ over all datasets with the same format or schema as $D$. For instance, $P$ could simply be uniform over all datasets with the same schema as $D$, but any $P$ suffices in the argument that follows. Let us assume that the private dataset $D$ is drawn according to this prior (denoted $D \sim P$), and we are given some queries $Q$ as well as their numerical values on $D$, denoted $Q(D)$. Let us assume for the moment\footnote{Shortly we shall discuss the realism of this assumption.} that if we initialize \rap~at a sample drawn from the prior $P$ and run it once, then the resulting reconstructed dataset $D'$ is actually a sample from the posterior distribution given the queries and their computed values: $D' \sim P|\langle Q,Q(D) \rangle$.
How could we use the ability to sample such datasets $D'$ from the posterior 
to estimate the probability that particular points $x$ are elements of $D$?

Let $\chi(D,D')$ be any random variable determined by the draws $D \sim P, D' \sim P|\langle Q,Q(D)\rangle$ --- for instance, a natural $\chi(D,D')$ for our purposes would take value equal to 1 if both $D$ and $D'$
contain some particular row $r$, and 0 otherwise. 
The attacker is interested in the expectation
\begin{equation} \label{firstexp}
    \mathbb{E}_{D \sim P, D' \sim P|\langle Q,Q(D)\rangle}[\chi(D,D')]
\end{equation}
which in the example above is simply the probability that both $D$ and $D'$ contain the row r. The difficulty is that although given $Q(D)$, we have assumed that we can take samples $D' \sim P|\langle Q,Q(D)\rangle$, we cannot evaluate the predicate $\chi(D,D')$ because we do not have access to the true dataset $D$ from which the statistics $Q(D)$ were computed. 

However, it is not hard to derive that this expectation is identical to:
\begin{equation} \label{secondexp}
    \mathbb{E}_{D \sim P} \left[\mathbb{E}_{ \tilde{D} \sim P|\langle Q,Q(D)\rangle,  D' \sim P|\langle Q,Q(D)\rangle}\left[\chi(\tilde{D},D')\right]\right]
\end{equation}
In other words, rather than computing $\chi(D, D')$ we can instead compute $\chi(\tilde D, D')$ where $\tilde D$ and $D'$ are both independent samples from the posterior distribution $P|\langle Q,Q(D)\rangle$ --- i.e. under our assumption, two reconstructions that result from running $\rap$~twice with fresh randomness. The reason for this equivalence is that in the two expectations above, the joint distributions of $\langle D, Q(D), D' \rangle$ and $\langle \tilde{D}, Q(D), D' \rangle$ are identical, since $D$ and $D'$ are conditionally independent given $Q(D)$, and both are distributed according to $P|\langle Q,Q(D)\rangle$.
In other words, if we wish to estimate the expectation [\ref{firstexp}], we can do so by instead estimating the expectation [\ref{secondexp}], which involves evaluating the predicate only on datasets drawn from the posterior rather than the prior. Concretely, rows that are more likely to appear in two or more draws from the posterior are also more likely to appear in $D$ drawn from the prior and $D'$ from the posterior. 

The assumption above that \rap~samples from the posterior is in general unrealistic, and we have
no formal evidence to support it.
But to the extent that \rap~even approximates draws from the posterior given $Q(D)$,
this argument provides a plausible explanation for why the ranking $R(D^*)$ might approximately correspond to an ordering of data points by their probability of appearing in the true dataset $D$. In general sampling from a posterior in a space of high-dimensional datasets and queries is a computationally intractable problem  \cite{pmlr-v99-tosh19a}, but this does not rule out effective heuristics on real datasets, and we believe that this Bayesian argument provides at a minimum some insight about why methods such as ours work well in practice.

\section{Empirical Findings}

In this section we describe our primary experimental findings. Additional and more fine-grained results
are provided in the appendix, including plots of \matchrate~for all 50 states on the Census data.

\subsection{Datasets and Queries}\label{sec:datasetandqueries}

\subsubsection{U.S. Decennial Census} \label{sec:PPMF}

\paragraph{Dataset:}
We conduct experiments on subsets of synthetic U.S. Census microdata released by the Census Bureau during the development of the 2020 Census Disclosure Avoidance System (DAS).
This synthetic microdata was generated so that it has similar statistics when compared to the real 2010 Census microdata.
We use the 2020-05-27 vintage Privacy-protected Microdata File (PPMF) \cite{ppmf20200527vintage}.
In our experiments, we treat the PPMF as the ground truth microdata, even though it is synthetic, since the true microdata has never been released. Despite being generated with the intent to mimic statistics of the real Census microdata, we find that it has several statistical peculiarities\footnote{For example, the PPMF data has an unusually large number of data points that appear exactly 5 times in their block or tract. This statistical artifact persists even for small blocks: for example, the data has blocks of size 20 that consist of only four ``types'' of people, with each type appearing with cardinality 5. This has the effect of inflating the performance of the strongest of our baselines, since when splitting a dataset in half, a point that appears 5 times will constitute a match 93.75\% of the time (i.e. whenever it is not the case that all 5 copies of the point fall into the same half of the random split).}, so in this paper we emphasize the relative performance of our methods to baselines (all computed on the PPMF data), rather than the absolute numbers.

The 2020-05-27 vintage PPMF consists of 312,471,327 rows, each representing a (synthetic) response for one individual in the 2010 Deccenial Census.
The columns correspond to the following attributes: The location of the respondent's home (state, county, census tract, and census block), their housing type (either a housing unit, or one of 8 types of group quarters), their sex (male or female), their age (an integer in $\{0, \dots, 115\}$), their race (one of the 63 racial categories defined by the U.S. Office of Management and Budget Standards)\footnote{The 63 race categories correspond to any non-empty subset of the following: American Indian or Alaska Native, Asian, Black or African American, Native Hawaiian or Other Pacific Islander, White, and Other.}, and whether they have Hispanic or Latino origin or not.

We evaluate reconstruction attacks on subsets of the PPMF that contain all rows belonging to a given census tract or census block.
According to the U.S. Census Bureau, census tracts typically have between 1,200 and 8,000 people with an optimum size of 4000, and cover a contiguous area (although their geographic sizes vary widely depending on population density).
Each census tract is partitioned into up to 10,000 census blocks, which are typically small regions bounded by features such as roads, streams, and property lines.

In our tract-level experiments, we randomly select one tract from each state. In our block-level experiments, we select for each state the block closest in size to mean block size as well as the largest block. In addition, we select blocks closest in size to $M / C$, where $M$ is the maximum block size in the state and $C \in \{2, 4, 8, 16\}$. Thus in total, we evaluate on $50$ tracts and $300$ blocks.

\paragraph{Statistical Queries:} 
The U.S. Census Bureau publishes a collection of data tables containing statistics computed from the microdata at various levels of geographic granularity.
For example, some tables are published at the block level, meaning that they release a copy of that table for every census block in the U.S., while others are published at the census tract or county level.
Our experiments attempt to reconstruct the microdata belonging to census tracts and blocks based on statistics contained in the Census tables.

We use the same tables that the Census Bureau used in their internal reconstruction attack of the 2010 Census data~\cite{JASONPrivacyMethods2022}.
These are the following tables from the Census Summary File~1:\footnote{Summary File 1 has been renamed to the Demographic and Housing Characteristics File (DHC) for the 2020 U.S. Census. In all cases, we refer to tables and data products by their names used in the 2010 Census.}
\begin{description}[nosep,labelindent=0.3cm]
    \item[P1:] Total population,
    \item[P6:] Race (total races tallied),
    \item[P7:] Hispanic or Latino origin by race (total races tallied),
    \item[P9:] Hispanic or Latino and not Hispanic or Latino by race.
    \item[P11:] Hispanic or Latino, and Not Hispanic or Latino by race for the population 18 years and over,
    \item[P12:] Sex by age for selected age categories (roughly 5 year buckets),
    \item[P12 A-I:] Sex by age for selected age categories (iterated by race).
    \item[PCT12:] Sex by single year age.
    \item[PCT12 A-N:] Sex by single year age (iterated by race).
\end{description}
All of the P tables are released at the block level, while the PCT tables are released only at the census tract level.

Each table defines a collection of statistical queries that will be evaluated on the Census microdata.
For example, cell 3 of table P12 counts the number of male children under the age of 5.
Since P12 is a block-level table, cell 3 corresponds to one statistical query per census block.
Similarly, each cell in a tract-level table encodes one statistical query per census tract.
All of the statistical queries in the above tables can be encoded as follows: Given $k$ pairs $(\text{col}_i, S_i)_{i=1}^k$, where $\text{col}_i$ is a column name and $S_i$ is a subset of that column's domain, and either a census block or tract identifier,
count the number of microdata rows belonging to that tract or block for which $\text{col}_i \in S_i$ for all $i \in [k]$.\footnote{The Census tables report row counts, but in our experiments we convert counts to fractions by dividing by the population of the tract or block we are reconstructing.}
Thus in logical terms, queries are in Conjunctive Normal Form (CNF), meaning that they consist of a conjunction (logical AND) of clauses, with
each clause being a disjunction (logical OR) of allowed values for a column.

For example, cell 3 of table P12 encodes queries for each block with $\text{col}_1 = \text{Age}$, $S_1 = \{0, \dots, 4\}$, and $\text{col}_2 = \text{Sex}$, $S_2 = \{\text{Male}\}$.
When we perform tract-level reconstructions, we use queries defined by all of the above tables.
For block-level reconstructions, we use only the block-level tables (i.e., excluding tables PCT12 and PCT12 A-N).
In order to minimize the total number of queries, we omit several table cells that are either repeated or can be computed as a sum or difference of other table cells.

The statistical queries encoded by the Census data tables vary significantly in the value of $k$ (number of clauses in a conjunction) and the size of the sets (clauses) $S_i$. There are 2 cells with $k = 0$ (total population at the block and tract level), 27 cells with $k = 1$, 352 cells with $k = 2$, 1915 cells with $k = 3$, and 1259 cells with $k = 4$. 
The size of the sets $S_i$ range from 1 to 98.

To verify the correctness of our implementation of the statistical queries from the tables above, we compared the output of our implementation to tables released by the IPUMS National Historical Geographic Information System (NHGIS). 
For each vintage of the PPMF released by the U.S. Census Bureau, the NHGIS computes the census tables from that PPMF vintage.\footnote{The NHGIS tables constructed from each PPMF vintage are available at \href{https://www.nhgis.org/privacy-protected-2010-census-demonstration-data}{https://www.nhgis.org/privacy-protected-2010-census-demonstration-data}.}
We compared our implementations of queries from tables P1, P6, P7, P9, P11, P12, and P12 A-I on all census blocks in the United States and Puerto Rico and found no discrepancies.
Unfortunately, the PCT12 and PCT12 A-N tables were not included in the NHGIS tabulations for the 2020-05-27 vintage PPMF, so we were unable to verify our implementation of these queries (but their structure is very similar to the block-level queries).

\subsubsection{American Community Survey (ACS)}
\paragraph{Dataset}
We conduct additional experiments on a suite of datasets derived from US Census, introduced in \cite{ding2021retiring}.\footnote{The Folktables package comes with MIT license, and terms of service of the ACS data can be found at \href{https://www.census.gov/data/developers/about/terms-of-service.html}{https://www.census.gov/data/developers/about/terms-of-service.html}.}
The Folktables package defines datasets for each of 50 states and various tasks. 
Each task consists of a subset of columns\footnote{A detailed list of the attributes can be found in the \ifarxiv Appendix \else SI \fi (Table \ref{tab:folktable_columns}). Note that we discretize numerical columns into 10 equal-sized bins.} from the American Community Survey (ACS) corpus. These datasets provide a diverse and extensive collection of datasets helpful in experimenting with practical algorithms. 
We use the five largest states (California, New York, Texas, Florida, and Pennsylvania) which together with the three tasks (employment, coverage, mobility) constitute 15 datasets. Our experiments therefore seek to reconstruct individuals at the state-level. 
Compared to datasets derived from the Census Bureau's May 2020 Demonstration Data Product (PPMF), based on the 2010 Census, the Folktables ACS datasets contain many more attributes (see Table \ref{tab:folktables}), helping us demonstrate how our reconstruction attack scales up to higher dimensional datasets.

We note that while the datasets distributed by the Folktables package are derived from the ACS microdata, the package was designed for evaluating machine learning algorithms, and there exist many differences from the actual 1-year and 5-year statistical tables released by the Census Bureau each year. As mentioned above, each task only contains a subset of features collected on the ACS questionnaire and released in the 1-year Public Use Microdata Sample (PUMS). Moreover, survey responses are collected at both the household and person-level, but Folktables treats records only at the person-level. Lastly, in the ACS PUMS, each survey response is assigned a sampling weight, which can then be used to calculate weighted statistics (e.g., estimated population sizes and income percentiles) that estimate population-level statistics. Folktables ignores these weights, and so the statistics we calculate and use for experiments are unweighted tabulations. Folktables also ignores the replicate weights on the ACS PUMS that the Census Bureau recommends users implement to generate measures of uncertainty associated with the weighted statistics.

\paragraph{Statistical Queries} For each ACS dataset we compute a set of  $k$-way marginal statistics. A  marginal query counts the number of people in a dataset whose features match a given value. An example of a $2$-way marginal query  is: "How many people are \emph{female} and and have \emph{income greater than 50K}". The formal definition is as follows:
 \begin{definition}[$k$-way Marginal Queries]\label{def:marginals} Let $\cX=\prod_{i\in[d]}\cX_i$ be a discrete data domain with $d$ features, where $\cX_i$ is the domain of the $i$-th feature.
A $k$-way \textit{marginal query} is defined by a set of $k$  features $S\subseteq[d]$, together with a target value $v\in \prod_{i\in S} \cX_i$ for each feature in $S$.  Given such a pair  $(S, v)$, let $\cX(S, v)=\{x \in \cX : x_i = v_i \forall i\in S \}$ denote the set of points in $\cX$ that match $v$ on each feature $i\in S$.   Then consider the function $\phi_
{S, v}$ defined as $\phi_{S,v}(x) = \mathbbm{1}\{x\in \cX(S, v)\}$,  where $\mathbbm{1}$ is the indicator function.
The corresponding $k$-way marginal query is the statistical query defined as 
$$q_{S,v}(D) = \frac{1}{|D|}\sum_{x\in D} \phi_{S, v} (x)$$
for any dataset $D$.
\end{definition}

We explore the efficacy of our reconstruction attack on ACS datasets when all $2$-way or $3$-way marginal queries are released.

\begin{table}[htbp]
    \centering
    \caption{For each Folktables task, we list the number attributes, total dimension of such attributes, and the number of all 2 and 3-way marginal queries.}
    \begin{tabular}{l|rrrr}
    \toprule
           Task &  \# Attr &  Dim &  \# 2-way &  \# 3-way \\
    \midrule
     employment &      16 &  108 &     5154 &   144910 \\
       coverage &      18 &  107 &     5160 &   149848 \\
       mobility &      21 &  141 &     9137 &   362309 \\
    \bottomrule
    \end{tabular}
    \label{tab:folktables}
\end{table}

\subsection{Baselines}\label{sec:baseline}

In isolation, the \matchrate~of \rapattack~described in the previous section cannot provide enough information to indicate a privacy breach. If the dataset distribution is very low entropy, and we know the distribution, then we might expect to obtain a high \matchrate~simply by randomly guessing rows that are likely under the data distribution.  Therefore, we would like to compare the \matchrate~of our attack to the \matchrate~of baselines of various strengths corresponding to increasingly precise knowledge of the data distribution. 

Given a baseline distribution $P$, we consider a \matchrate~baseline that results from ordering the rows of $\cX$ according to their likelihood of appearing in a randomly sampled dataset $D\sim P$. In practice, the domain size $|\cX|$ is often too large to enumerate; an alternative in this case is to sample a large collection of rows $X \sim P$ and then compare to the confidence set $R(X)$---i.e. the ranking that results from ordering rows by their likelihood in the empirical distribution over $X$, sampled from $P$.

We compare to different baselines corresponding to a set of increasingly informed prior distributions. First, in order to simulate a prior that is identical  to the distribution from which the private dataset is sampled, we randomly partition the real dataset into two halves $D$ and $D_{holdout}$. We treat $D$ as the private dataset which we compute statistical queries on and seek to reconstruct rows from, while $D_{holdout}$ is used to produce a baseline confidence set $R(D_{\textrm{holdout}})$. Here, by construction, $D$ and $D_{holdout}$ are identically distributed, which allows us to compare to the very strong baseline of the ``real'' sampling distribution for real datasets. Of course, as a synthetic construction originating from the real data, $R(D_{holdout})$ should generally be viewed as an unrealistically strong benchmark. 

We also compare to a natural hierarchy of benchmarks that correspond to fixing a prior based on knowledge of Census data at different levels of granularity.  U.S. Census data is organized according to geographic entities that have a hierarchical structure. We consider a natural hierarchy of prior distributions in which a lower level in the hierarchy is more informative than higher levels. For example, for block-level reconstruction, we consider benchmarks defined by sampled rows from the tract, county, and state ($D_{\textrm{tract}}$, $D_{\textrm{county}}$, $D_{\textrm{state}}$) that each block is contained in, as well as the benchmark defined by samples from all rows in the dataset ($D_{\textrm{national}}$). We note that in block level reconstruction experiments, $D_{\textrm{holdout}}$ corresponds to a block-level prior, and so we refer to this set of rows as $D_{\textrm{block}}$ in Section \ref{sec:results}. Similarly for tract-level reconstruction experiments, $D_{\textrm{holdout}}$ is referred to as $D_{\textrm{tract}}$.

As we describe in more detail in Section \ref{sec:results}, we run reconstruction of Census tracts both with and without the attribute corresponding to the block each individual resides in. For the setting in which the block attribute is included, the county, state, and national baselines are at an extreme disadvantage, since the majority of individuals in $D_{\textrm{county}}$, $D_{\textrm{state}}$, and $D_{\textrm{national}}$ reside in a tract different than those found in $D$---and so necessarily have different block values. To compensate for this disadvantage (otherwise crippling to the baselines), in these cases we strengthen the baselines and instead populate the block attribute according to the distribution of blocks found in $D$. For example, the state-level baseline can be interpreted as a prior in which the distribution of blocks follows that of $D$ and the distribution of the remaining attributes follows that of $D_{\textrm{state}}$.

\subsection{Results}\label{sec:results}

Our primary reconstruction rate visualization technique is as follows.

Recall both \rapattack~and our baselines each output some confidence set $R$. Therefore, for both \rapattack~and our baselines, we plot $\matchrate_{D,k}(R)$ against $k$, or in other words, the fraction of candidates of rank $k$ or higher that exactly match some row in $D$.
Because the many datasets on which we run our reconstruction attack vary considerably in size, and in some of our plots we average our results over many datasets, in the ensuing plots we express rank $k$ as a fraction of the number of unique rows in $D$, which we denote as $u$. In other words, the $x$-axis measures ${k}/{u}$. 
This allows us to average results across different samples of data (e.g., different geographic entities for both Census and ACS experiments) on a common scale for the $x$-axis.\footnote{Let $\hat{u}$ be the number of unique rows in $D_{holdout}$. In cases where $\hat{u}$ is smaller than the number of unique rows in $D$, we instead set $u = \hat{u}$. Otherwise, our baseline derived from $D_{holdout}$ would be penalized for outputting a candidate list of size smaller than $u$.}

\begin{figure}[htbp]
    \centering
    \ifarxiv
    \includegraphics[width=0.80\columnwidth]{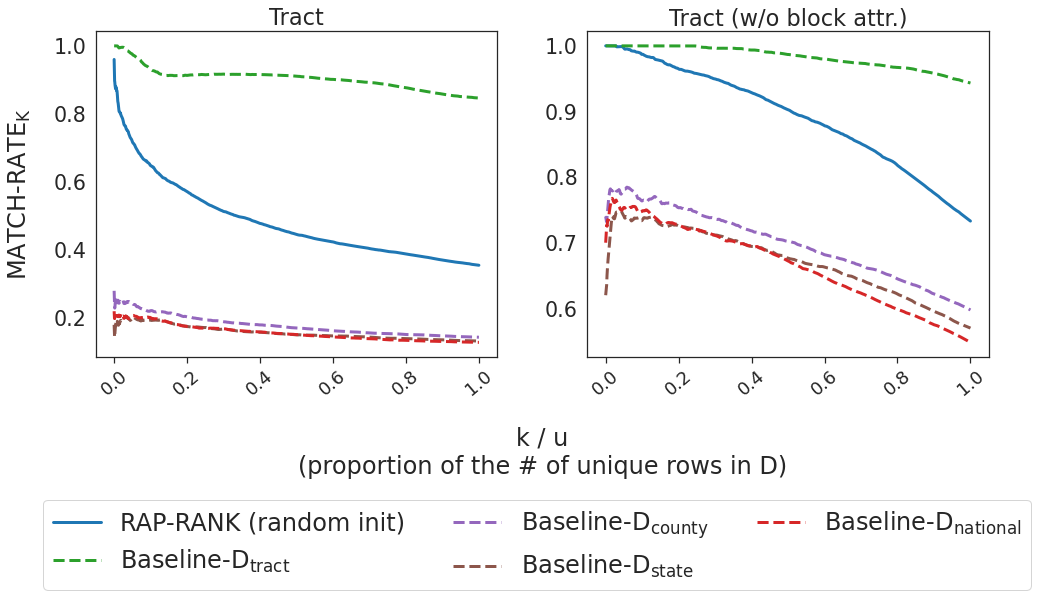}
    \else
    \includegraphics[width=\columnwidth]{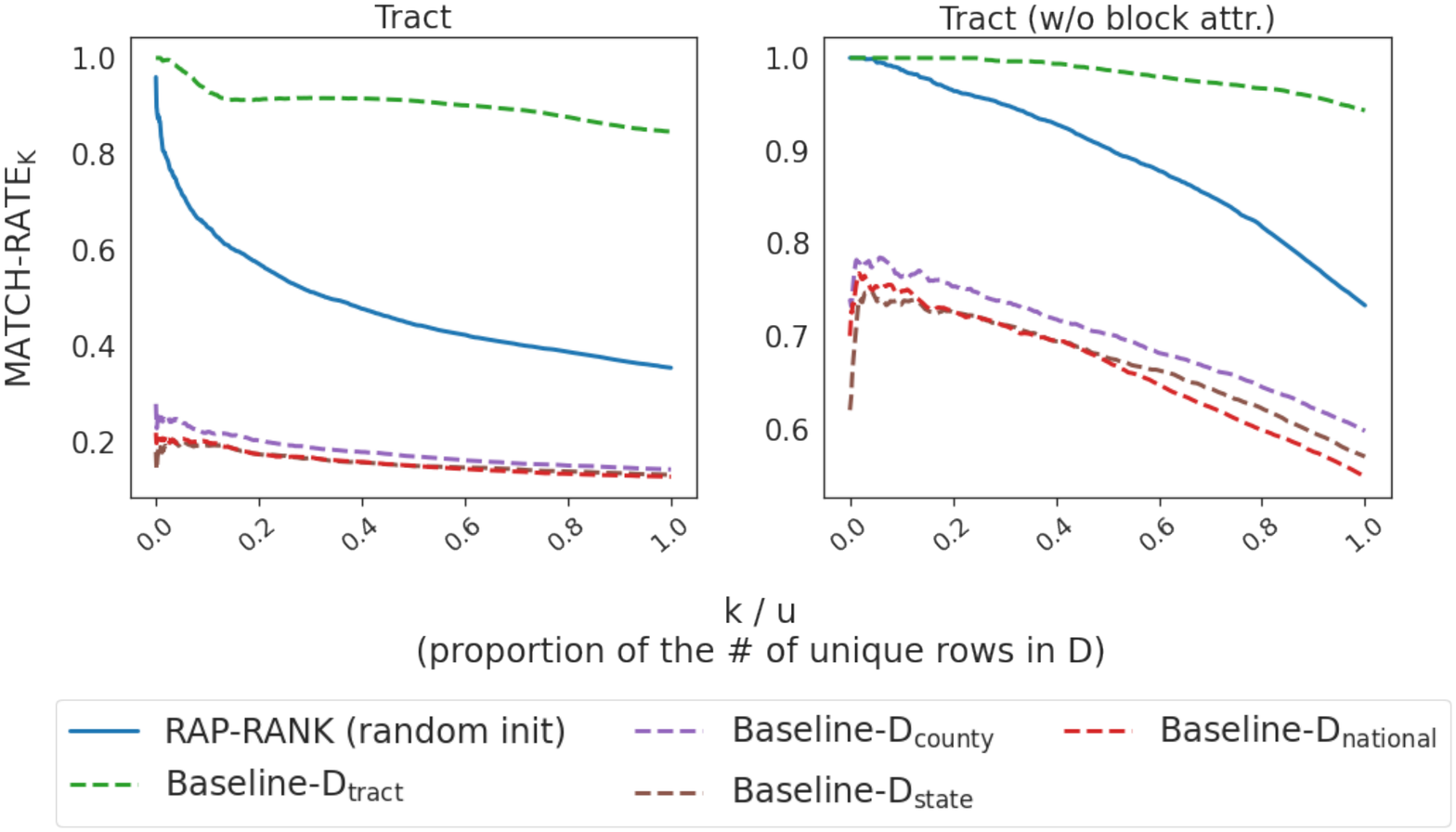}
    \fi
    \caption{The panel on the left plots the \matchrate~of \rapattack~and our various baselines on a tract level reconstruction when we use the BLOCK attribute. The panel on the right plots the \matchrate~of \rapattack~and our various baselines when the BLOCK attribute has been removed. Both panels show the average performance of \rapattack~and the baselines averaged over a randomly selected tract from each of the 50 US states. In both cases, \rapattack~is initialized uniformly at random (i.e. we have not initialized at a baseline distribution).}
    \label{fig-main:census_avg_tract}
\end{figure}

\begin{figure}[htbp]
    \centering
    \ifarxiv
    \includegraphics[width=0.6\columnwidth]{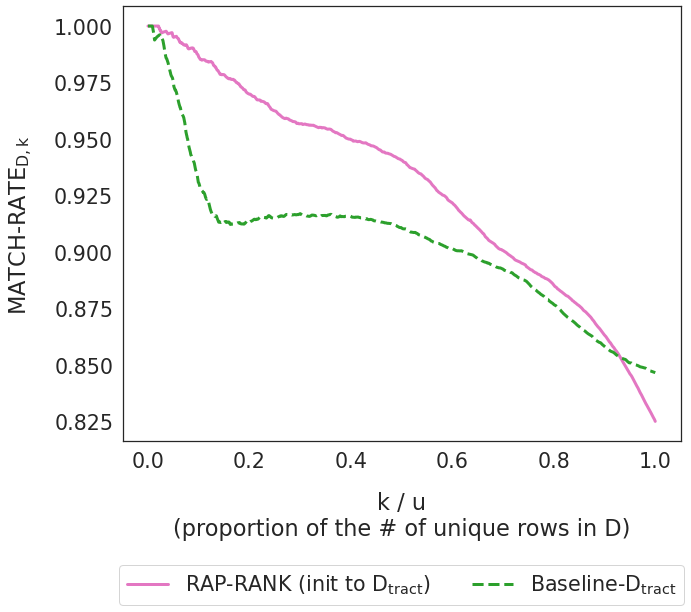}
    \else
    \includegraphics[width=0.7\columnwidth]{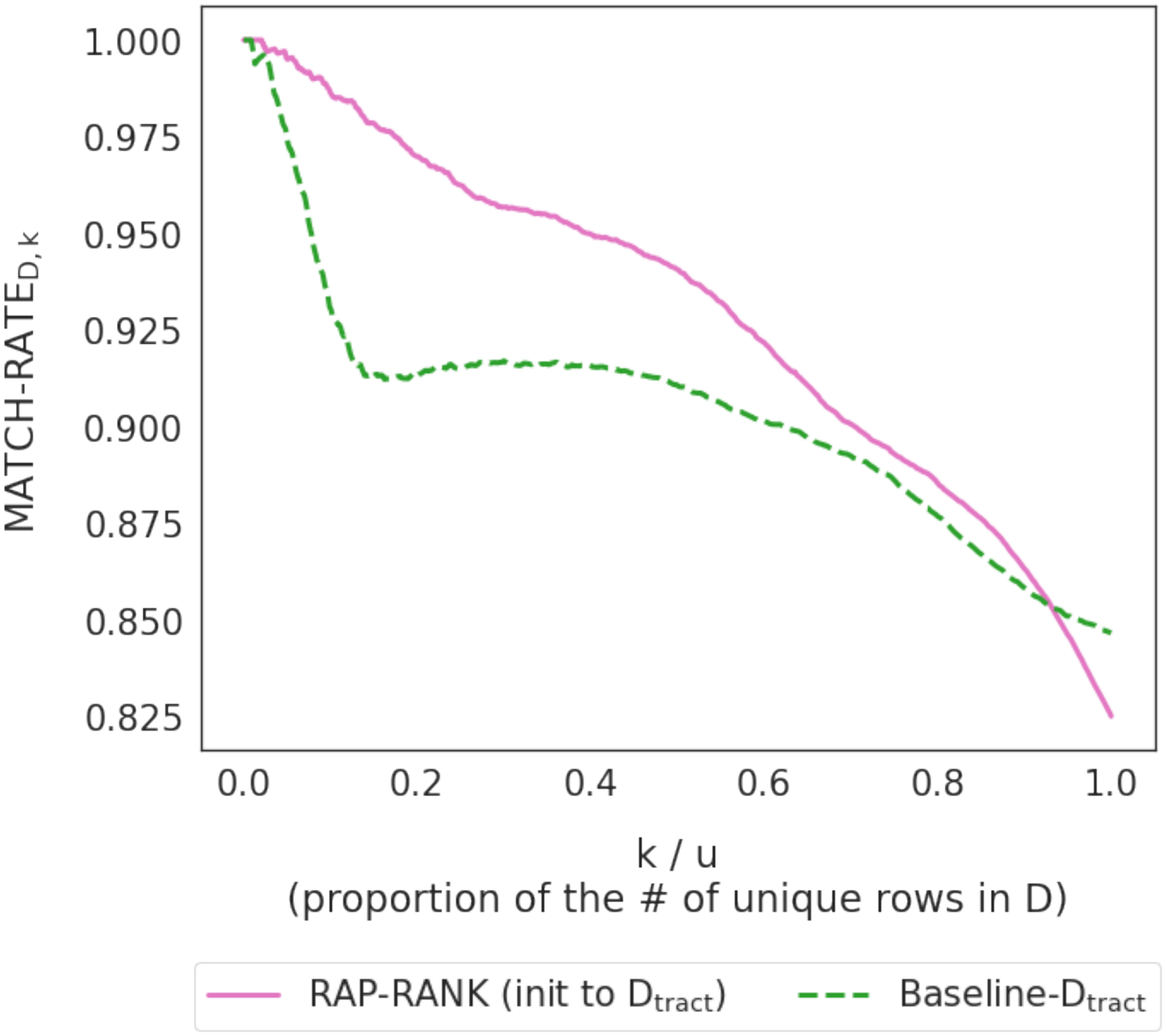}
    \fi
    \caption{Initializing \rapattack~at the tract baseline significantly improves its performance, leading it to out-perform the tract baseline. Here the BLOCK attribute is included and must be reconstructed to constitute a match.}
    \label{fig-main:census_avg_tract_init}
\end{figure}

\begin{figure}[htbp]
    \centering
    \ifarxiv
    \includegraphics[width=0.9\columnwidth]{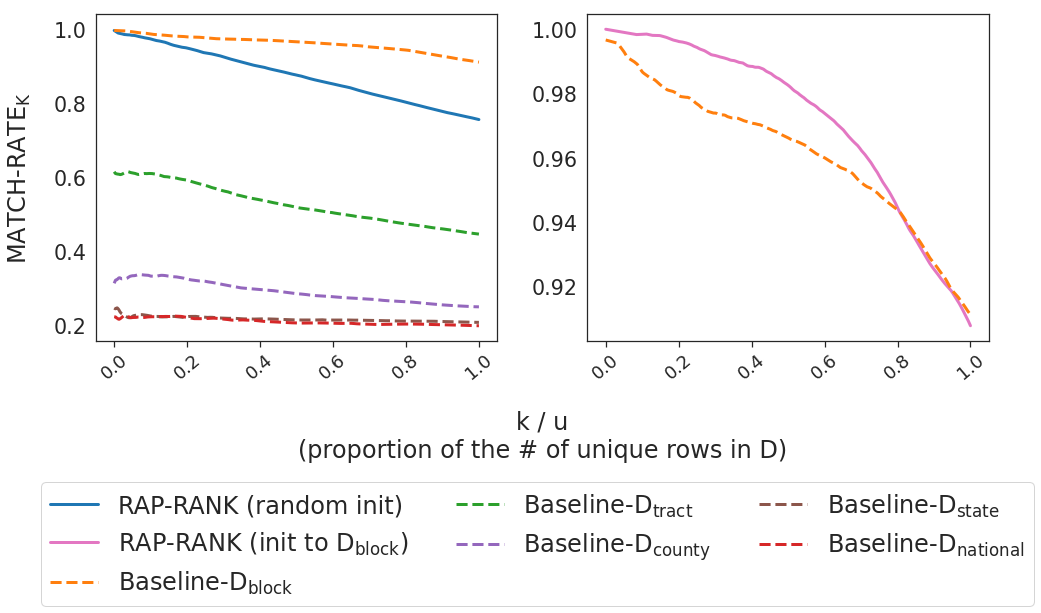}
    \else
    \includegraphics[width=\columnwidth]{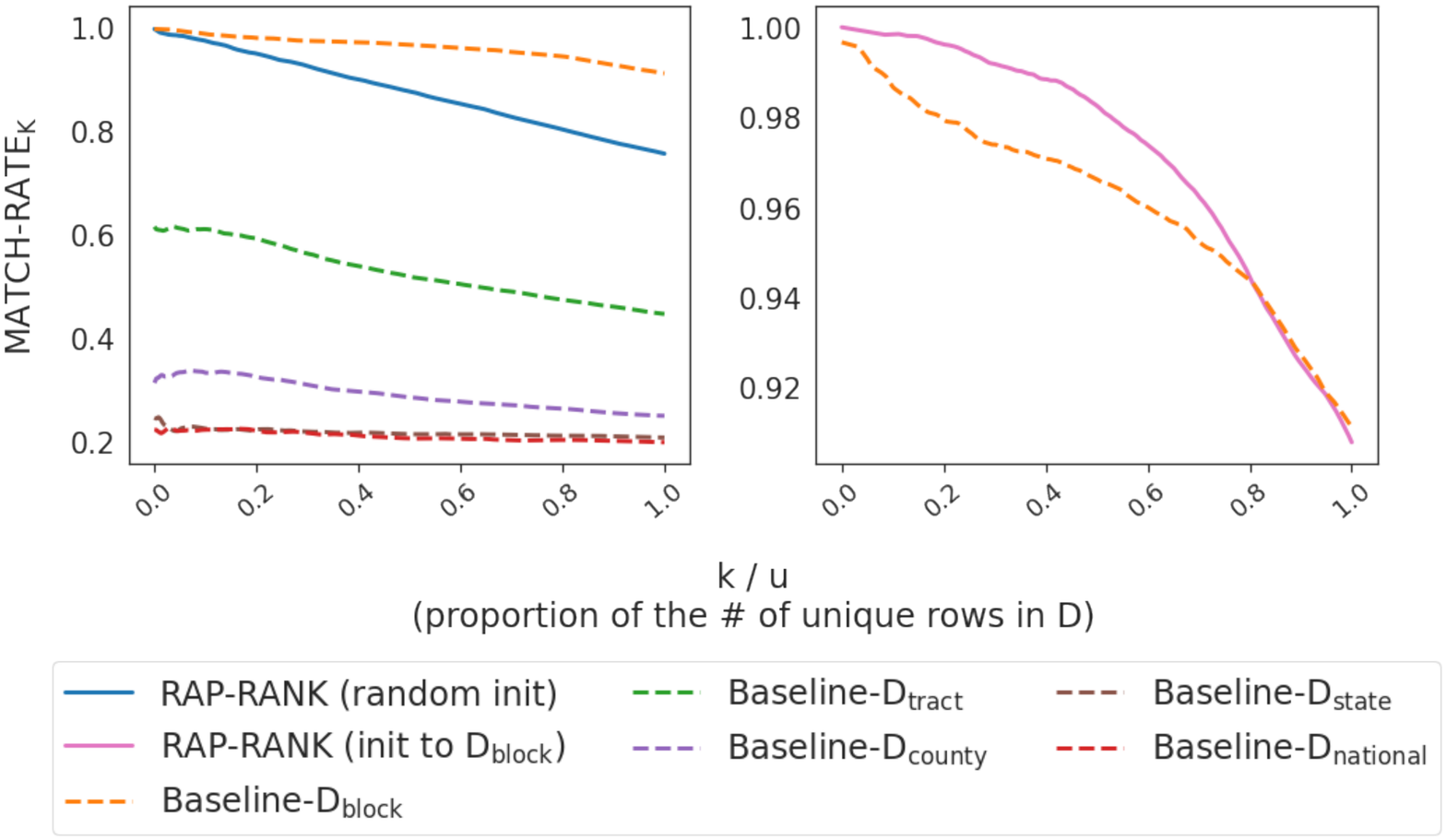}
    \fi
    \caption{The panel on the left plots the \matchrate~of \rapattack~and our various baselines on a block level reconstruction, when \rapattack~is initialized to a uniformly random dataset. The panel on the right shows the performance of \rapattack~when it is initialized to $D_{block}$, and compares its performance to $D_{block}$. }
    \label{fig-main:census_avg_block}
\end{figure}

\begin{figure*}[htbp]
    \centering
    \includegraphics[width=0.8\textwidth]{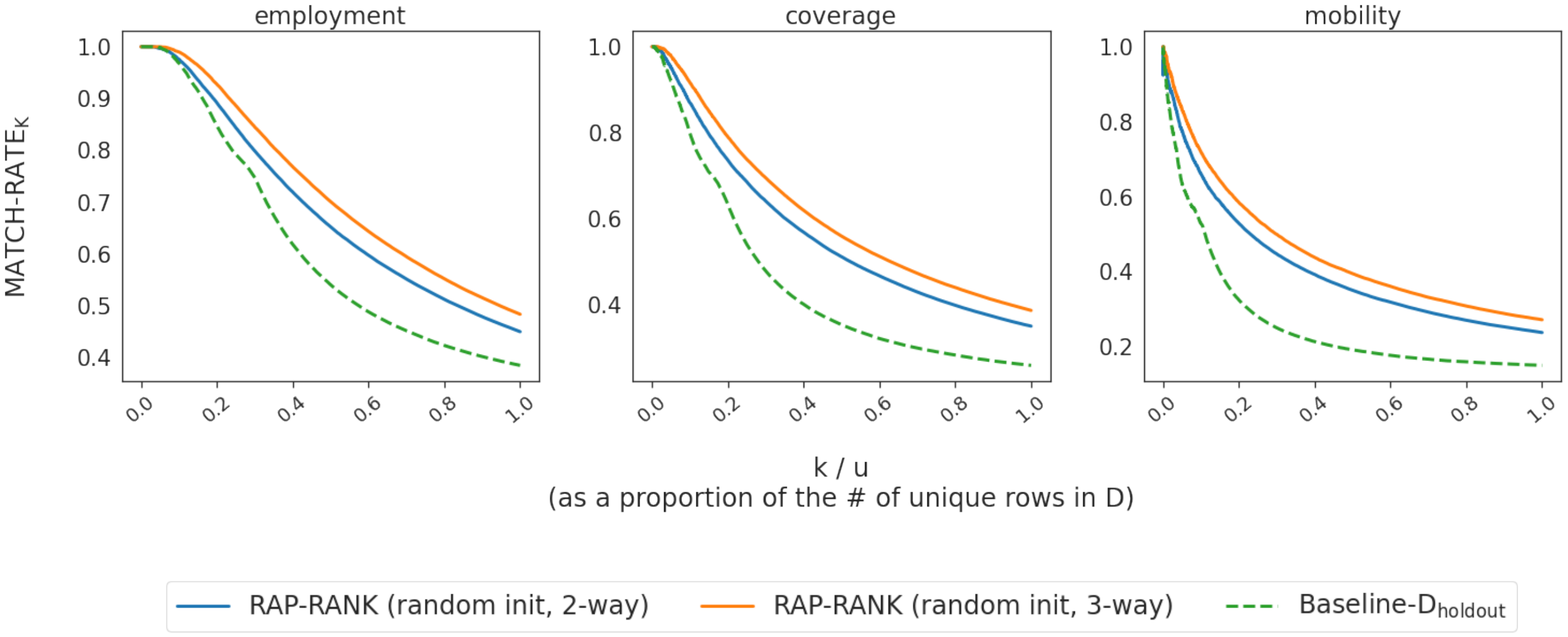}
    \caption{We select state-task combinations from the folktables dataset, comparing top $k$ candidate \matchrate~ of \rapattack~ against the baseline, which is derived from the holdout split. For each task, we average results over the five largest states (by population) in the United States (i.e., California, New York, Texas, Florida, and Pennsylvania). We initialize \rapattack~ randomly, showing results where $Q$ is all $2$ and $3$-way marginals. }
    \label{fig-main:folktables_avg}
\end{figure*}

In our first set of experiments, we randomly select a tract from each state, which forms the private dataset $D$ from which we compute the Census query-answer pairs. 
We run {\rapattack} using these queries and starting from a uniformly random initialization,\footnote{We shortly describe a natural and realistic alternative initialization
scheme that improves performance considerably.} and plot the match rate as a function of $k/u$. We similarly plot the match rate of each of our baselines. 
In the left panel of Figure~\ref{fig-main:census_avg_tract}, we plot the reconstruction rates after averaging across the selected tract from all 50 states.
(See Figures \ref{fig-appx:census_tract_all1}, \ref{fig-appx:census_tract_all2}, \ref{fig-appx:census_tract_ib_all1}, and \ref{fig-appx:census_tract_ib_all2} in the \ifarxiv appendix \else SI \fi for the state-by-state plots that comprise this average.)

As expected, in general at higher ranks (lower $x$-axis values) the reconstruction rates are reasonably high and then fall
at lower ranks.  The left panel shows that the \rapattack~reconstruction rates are considerably higher than all but the
strongest baseline --- resampling at the tract level --- which is much higher still. Recall that since this is a tract-level reconstruction, $D_{tract}$ here is in fact $D_{holdout}$ --- i.e. the very strong artificial benchmark constructed from the dataset we are attacking itself. We see that the other baselines---$D_{county}, D_{state}$, and $D_{national}$ perform quite poorly. This is partially an artifact of requiring that they reconstruct the BLOCK attribute. Since blocks appearing within a tract appear in no other tracts, the non-tract baselines have a poor chance of reconstruction since they are sampling at a coarser geographic level. Recall that we have strengthened these baselines by letting the BLOCK attribute be distributed according to the empirical distribution of blocks in the true dataset $D$, but still, these baselines are at a disadvantage because they have lost the correlation between the BLOCK attribute and all other features. 
Therefore in the right panel of Figure \ref{fig-main:census_avg_tract}, we reproduce the same experiment in which we have dropped the BLOCK attribute. This makes the reconstruction task easier and improves the performance of \rapattack~as well as all of the baselines. The most dramatic increase is in the performance of the $D_{county}, D_{state}$, and $D_{national}$ baselines, but we also see that \rapattack~now performs relatively better compared to the $D_{tract}$/$D_{holdout}$ baseline.  \rapattack~now has reconstruction rates above 0.9 up to $k/u \approx 0.5$.
These results establish that \rapattack~can perfectly reconstruct rows well beyond what sampling access alone permits except at the
most local level. In other words, \rapattack~is far from 
simply ``getting lucky'' --- its optimization process is deliberately and effectively exploiting the actual query-answer pairs,
not simply benefiting from having data similarly distributed to the private dataset. Nevertheless, the ordering of the baselines and of \rapattack~is unchanged --- i.e. \rapattack~outperforms all of the baselines \emph{except} for the artificial $D_{tract}$ baseline.

We next observe that there is an asymmetry in our experiments that treats \rapattack~in what could be considered an unfair manner: we assert that there are strong ``baseline'' distributions that are related to the data we are trying to reconstruct, and yet we have initialized our attack \rapattack~at a uniformly random dataset, without giving it the benefit of this knowledge. If indeed these baseline distributions are public knowledge, then an attacker could make use of them as well. Thus our next set of experiments consists of initializing \rapattack~at the baseline that we are comparing it to, and see that this causes it to significantly outperform all baseline---including $D_{tract}$ (which we recall is the strong baseline $D_{holdout}$), even with the BLOCK attribute. In other words, if we view the baseline as a public prior distribution, then giving \rapattack~access to it leads to the ability to significantly improve over it. 

In  Figure~\ref{fig-main:census_avg_tract_init}, we show results averaged across randomly chosen
tracts for all 50 states in which we have now initialized to the tract baseline and compare to that sampling baseline, (once again including the BLOCK feature).
The results are clear: when we level the playing field by seeding \rapattack~with knowledge of the tract baseline distribution,
it now outperforms the tract baseline. We can interpret the area between the two curves in Figure~\ref{fig-main:census_avg_tract_init}
as a measure of the additional reconstruction risk introduced by \rapattack~on the query-answer pairs, beyond the baseline
risk of tract sampling. 

In Figure~\ref{fig-main:census_avg_block}, we show that \rapattack~remains an effective reconstruction attack even at the
most fine-grained geographic level, which corresponds to Census blocks. The left panel again shows \matchrate~for \rapattack~initialized randomly, and compared to all the sampling baselines. Here we again see the same qualitative performance --- even with random initialization, \rapattack~outperforms all of the sampling baselines except for constructed $D_{block}$ (which we recall in this case is the artificially constructed  $D_{holdout}$). The right panel shows results when we initialize \rapattack~at $D_{block}$. In this case we again see that initializing at the benchmark distribution causes $\rapattack$ to significantly outperform the benchmark. This figure is again averaging over attacks on blocks from all 50 states. (See Figures \ref{fig-appx:census_block_all1} and \ref{fig-appx:census_block_all2} in the \ifarxiv appendix \else  SI \fi for the state-by-state plots that comprise this average.)

We conclude by briefly describing a second set of experiments on three datasets from the ACS Folktable package, corresponding
to the employment, coverage and mobility tasks. We consider these alternate datasets both to show the generality of our methods beyond
decennial Census data (in particular, the ACS Folktables datasets have much higher dimensionality than the decennial Census data), and in order to do a controlled comparison of queries of differing power (as opposed to the fixed set of queries provided for
the decennial Census data).

In Figure~\ref{fig-main:folktables_avg}, we show the reconstruction rates obtained by letting the query set $Q$ be the sets of all
2-way and 3-way marginal queries on these three ACS datasets, and as in the Census data we compare to the very strong $D_{holdout}$ baseline. 
Two remarks are in order. First, despite the low complexity of these queries compared to Census queries --- 2-way and 3-way marginals reference only pairs and 
triples of columns, respectively --- both considerably outperform the $D_{holdout}$ baseline even when \rapattack~is initialized randomly, maintaining reconstruction rates well above 0.8
even at the lowest rank. This suggests that not only is aggregation insufficient
for privacy, neither is restriction to simple queries. In fact, on this dataset, and with these simple queries, our reconstruction attack performs even better---outperforming the strongest baseline even without the benefit of being initialized at that baseline. 

Second, the lift in performance in moving from 2-way marginals to 3-way marginals is large,
demonstrating the reconstructive power of even slightly more complex queries.

\section{Limitations and Conclusions}
We have shown the power of a new class of reconstruction attacks that can not only produce a candidate reconstructed dataset with a high intersection with the  true dataset, but also produce a ranking of rows that empirically corresponds to their likelihood of appearing in the true dataset. We have shown that from statistics that were actually released as part of the 2010 Decennial U.S. Census, it is possible to run our attack and that its \matchrate~ is high --- particularly at lower values of $k$, indicating high confidence reconstruction of a subset of the rows. Moreover, even with random initialization (equivalently, viewing \rapattack~as having an uninformative prior), \rapattack~outperforms all but the most stringent (artificial) benchmark that we construct. Finally, we can reliably outperform even the most stringent benchmark if we initialize \rapattack~at the benchmark distribution---consistently with the premise that if a distribution is publicly known (and so is sensible to consider as a public benchmark), then we should assume that attackers can make use of it as well. 

Nevertheless, our attack is not without limitations. First and foremost, our reconstructions of Census decennial data are far from recovering every row in the private data. The primary threat is that we can recover \emph{some fraction} of the rows with confidence. Moreover, our attack does not produce calibrated confidence scores. That is, we produce a ranking of rows $R$, but an attacker without access to the ground-truth would be unable to compute the \matchrate~as a function of $k$ as we do in our plots, and so would not know a-priori how much confidence to put in each reconstructed row. We highlight that developing an attack that can produce calibrated estimates of its match rate is a concrete technical problem whose solution  would improve our attack. Nevertheless, a ranking (known to be empirically correlated with \matchrate) is sufficient for an attacker to prioritize the rows of a reconstruction for some other external validation procedure or attack.



\ifarxiv
\subsection*{Acknowledgements}
We give warm thanks to Aloni Cohen and Kunal Talwar for their careful and detailed reviews, and
for making many valuable suggestions for improvements. This work was supported in part by NSF grants CCF-2217062 and FAI-2147212 and a grant from the Simons Foundation.
\else
\acknow{
We give warm thanks to Aloni Cohen and Kunal Talwar for their careful and detailed reviews, and
for making many valuable suggestions for improvements. This work was supported in part by NSF grants CCF-2217062 and FAI-2147212 and a grant from the Simons Foundation.
}
 \showacknow{} 
\fi

\ifarxiv
\bibliographystyle{alpha}
\else
\fi

\bibliography{pnas-sample}


\ifarxiv
\appendix
\section{Appendix}
\else
\section{Supporting Information}
\fi

\maketitle

We run \rapattack~with $K=100$ in all experiments. In our attack, each run of \rap~ sets $N'=1000$ and optimizes over query answers using ADAM with a learning rate of $0.1$. In Census experiments in which \rapattack is initialized to the baseline, $N'$ is instead the size of the $D_{holdout}$. In total, the number parameters of our attack is $K \times N' \times d$, where $d$ is the dimension size of the data domain.

In Table \ref{tab:folktable_columns}, we describe the columns used for each Folktables task found in our ACS experiments. 

\begin{table}[ht]
\centering
\caption{We detail below the Folktables columns we use for each task.}
\label{tab:folktable_columns}
\begin{tabular}{l | l}
    \toprule
    Task & Columns \\
    \midrule
    \multirow{8}{*}{employment}
    & AGEP (age), SCHL (educational attainment) \\
    & MAR (marital status), RELP (relationship) \\
    & DIS (disability recode), ESP (employment status of parents) \\
    & CIT (citizenship status), MIG (mobility status - lived here 1 year ago) \\
    & MIL (military status), ANC (ancestry recode) \\
    & NATIVITY (nativity), DEAR (hearing difficulty) \\
    & DEYE (vision difficulty), DREM (cognitive difficulty) \\
    & SEX (sex), RAC1P (recoded detailed race code) \\
    \midrule
    \multirow{9}{*}{coverage}
    & AGEP (age), SCHL (educational attainment) \\
    & MAR (marital status), SEX (sex) \\
    & DIS (disability recode), ESP (employment status of parents) \\
    & CIT (citizenship status), MIG (mobility status - lived here 1 year ago) \\
    & MIL (military status), ANC (ancestry recode) \\
    & NATIVITY (nativity), DEAR (hearing difficulty) \\
    & DEYE (vision difficulty), DREM (cognitive difficulty) \\
    & PINCP (Total person’s income), ESR (employment status recode) \\
    & FER (gave birth within the past 12 months), RAC1P (recoded detailed race code) \\
    \midrule
    \multirow{11}{*}{mobility}
    & AGEP (age), SCHL (educational attainment) \\
    & MAR (marital status), SEX (sex) \\
    & DIS (disability recode), ESP (employment status of parents) \\
    & CIT (citizenship status), MIL (military status) \\
    & ANC (ancestry recode), NATIVITY (nativity) \\
    & RELP (relationship), DEAR (hearing difficulty) \\
    & DEYE (vision difficulty), DREM (cognitive difficulty) \\
    & RAC1P (recoded detailed race code), GCL (grandparents living with grandchildren) \\
    & COW (class of worker), ESR (employment status recode) \\
    & WKHP (usual hours worked per week past 12 months), JWMNP (Travel time to work) \\
    & PINCP (Total person’s income) \\
    \bottomrule
\end{tabular}
\end{table}

\begin{figure*}[hb]
    \centering
    \includegraphics[width=\textwidth]{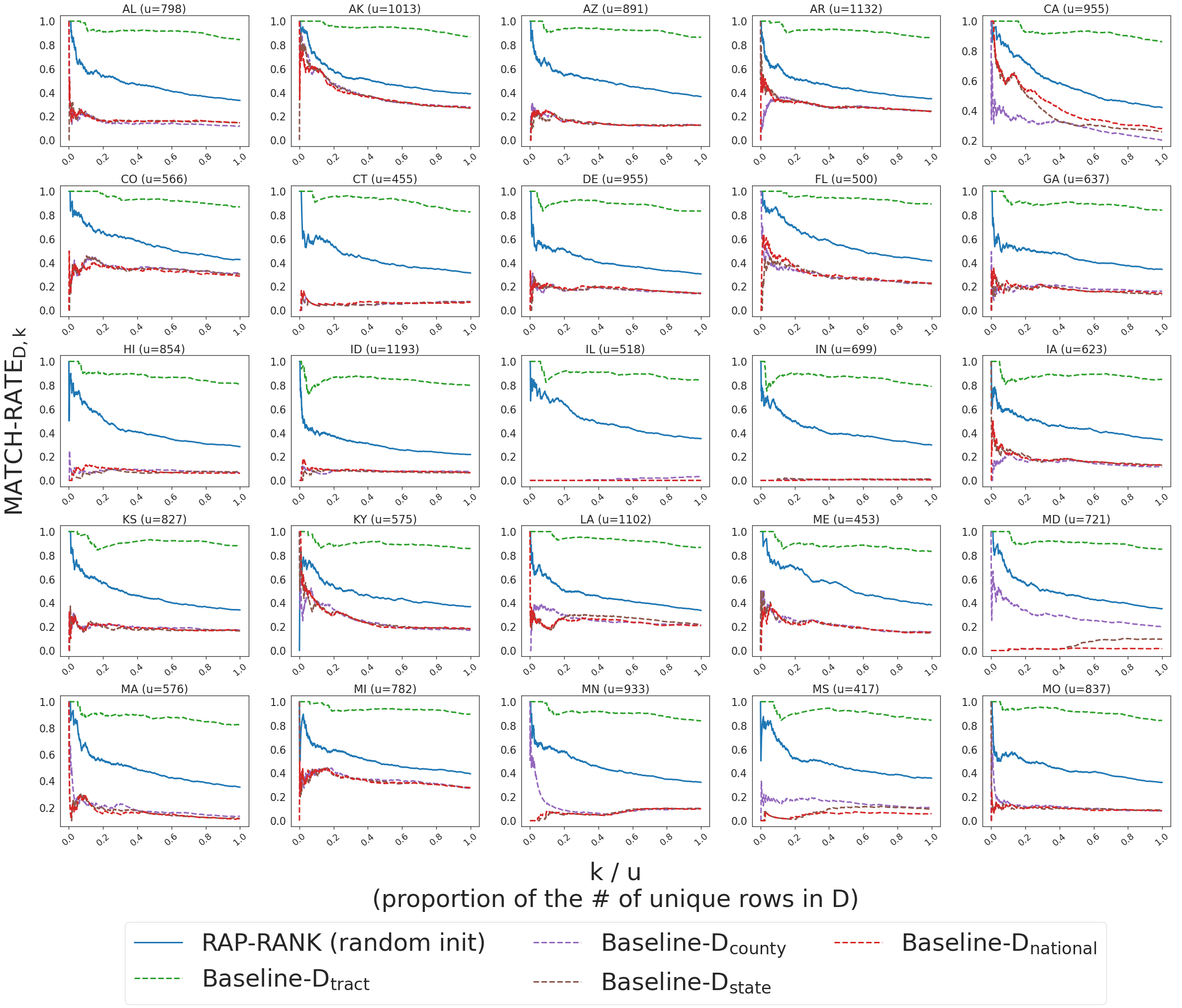}
    \caption{We plot \matchrate~of \rapattack~and our baselines on a \emph{tract}-level reconstruction with the BLOCK attribute \emph{included}. Subplots are labeled and ordered alphabetically by the state name.}
    \label{fig-appx:census_tract_all1}
\end{figure*}

\begin{figure*}
    \centering
    \includegraphics[width=\textwidth]{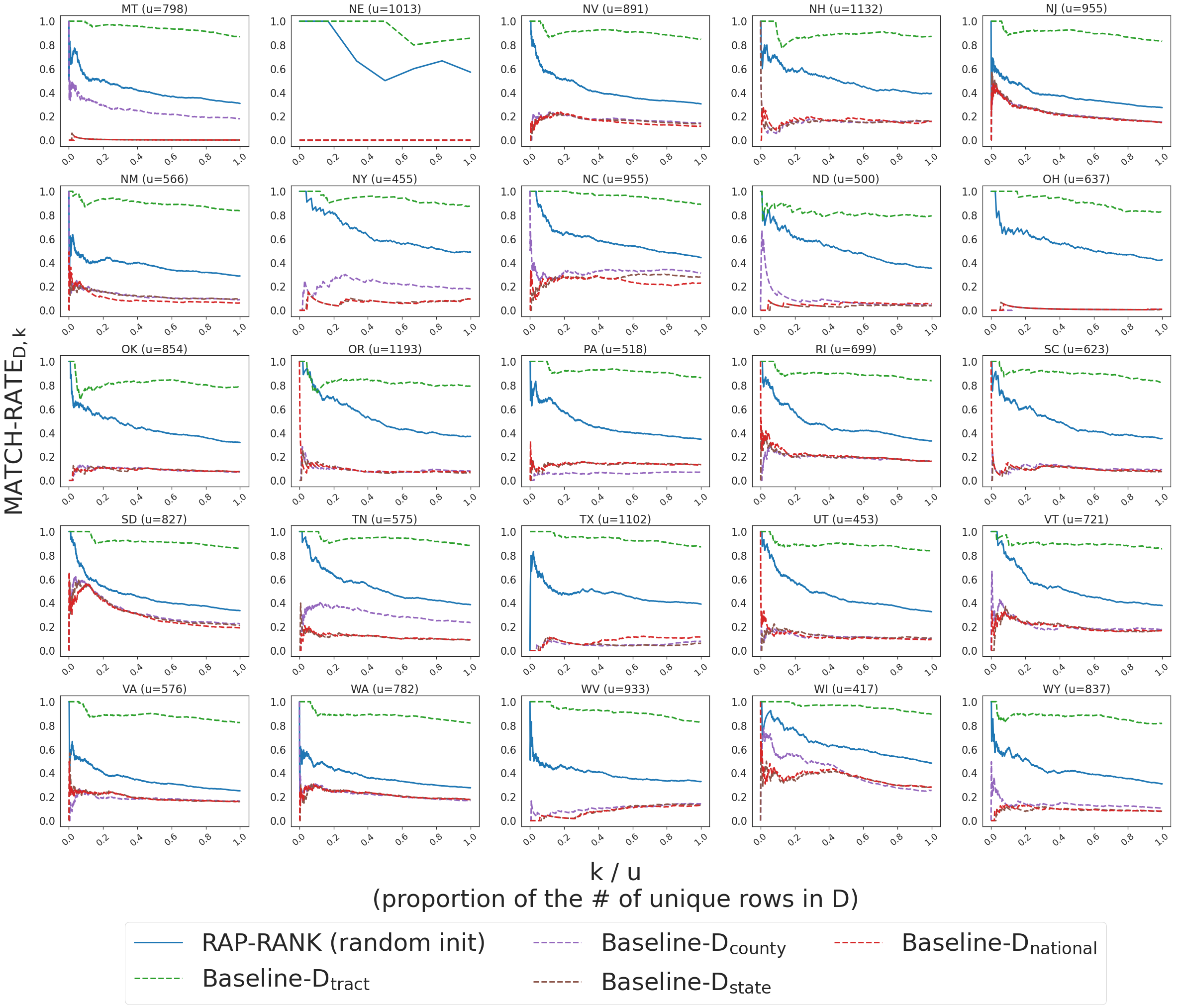}
    \caption{We plot \matchrate~of \rapattack~and our baselines on a \emph{tract}-level reconstruction with the BLOCK attribute \emph{included}. Subplots are labeled and ordered alphabetically by the state name.}
    \label{fig-appx:census_tract_all2}
\end{figure*}

\begin{figure*}
    \centering
    \includegraphics[width=\textwidth]{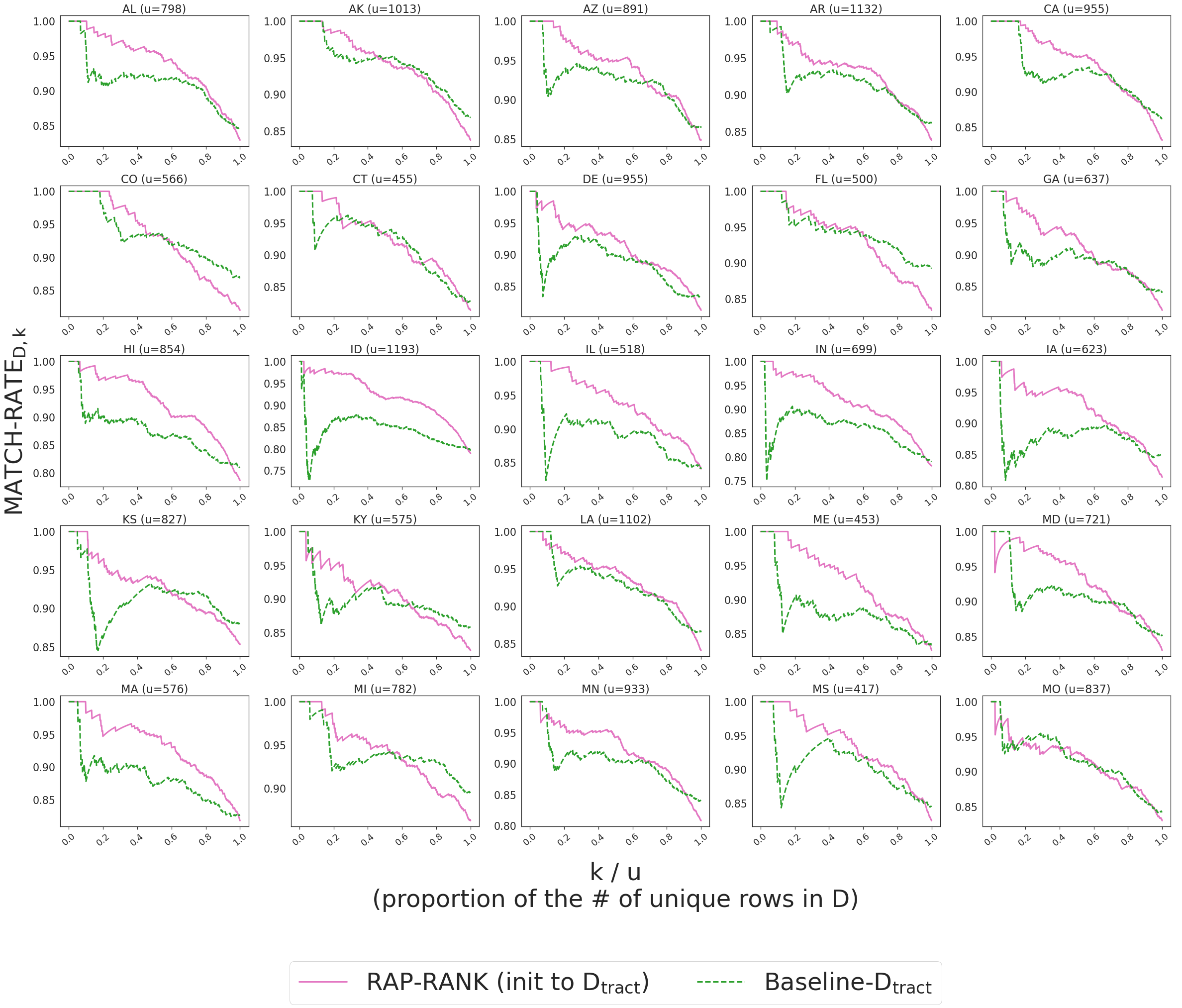}
    \caption{We plot \matchrate~of \rapattack~and our tract-level baseline on a \emph{tract}-level reconstruction with the BLOCK attribute \emph{included}. \rapattack~is initialized to the baseline. Subplots are labeled and ordered alphabetically by the state name.
    The sharp drops in the baseline match rates are explained by an anomaly in the raw PPMF microdata discussed in
    Section~\ref{sec:PPMF}, which is that in histograms
    of the number of unique rows with frequency $f$ in the overall tract, there is often a sharp spike at $f = 5$. Note that the probability
    that all $f$ copies of such a row fall into the baseline split is $1/2^f$, and that if this happens it will result in
    a non-match. For example, in the tract for AZ, there are 389 unique rows with $f = 5$ pre-split, while the counts for $f= 4$ and $f=6$ are
    88 and 27 respectively. The combination of this spike at $f=5$ and the fact that $1/32$ is relatively large results in many tracts
    having the observed dips in match rates. 
    }
    \label{fig-appx:census_tract_all_init1}
\end{figure*}

\begin{figure*}
    \centering
    \includegraphics[width=\textwidth]{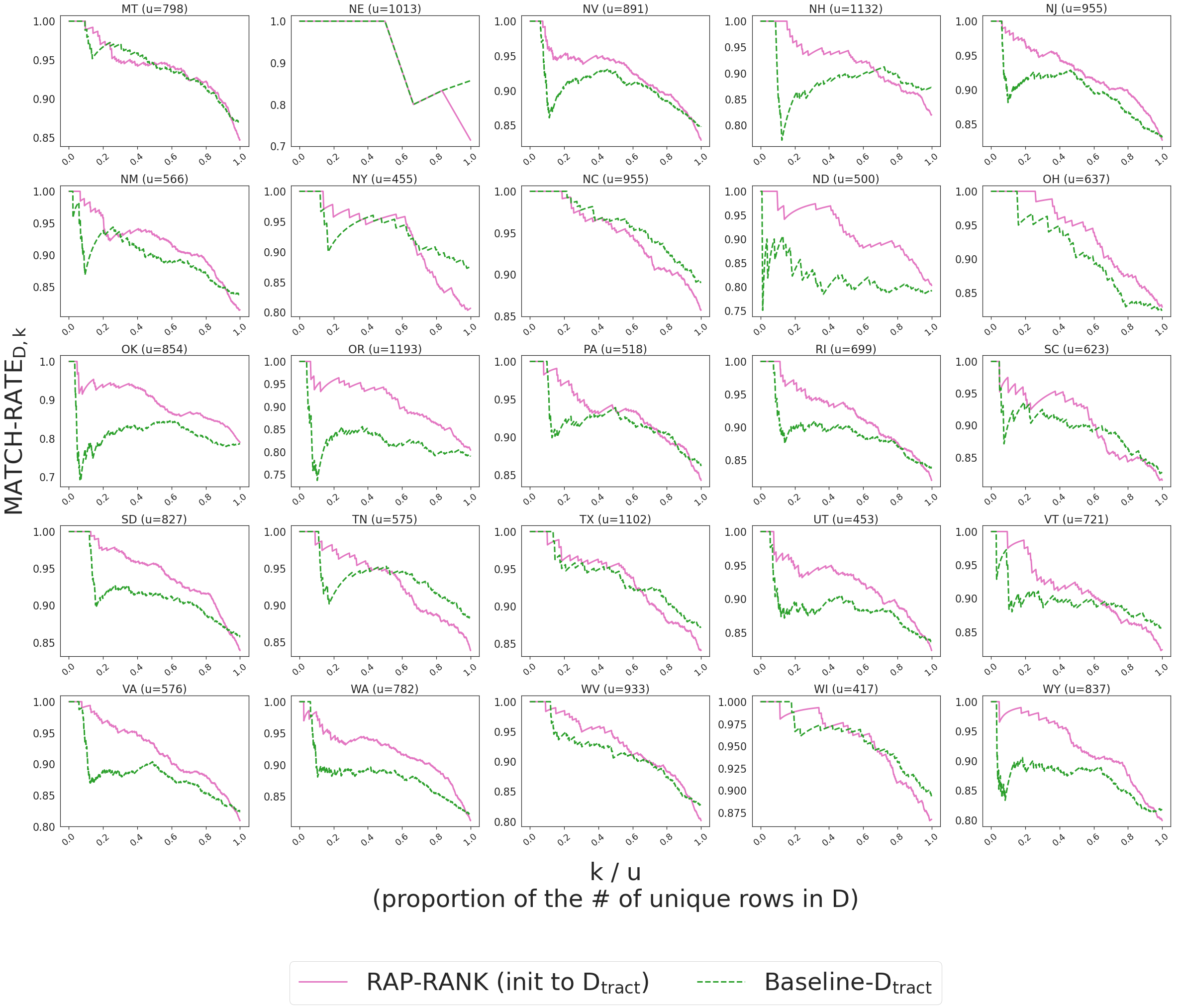}
    \caption{We plot \matchrate~of \rapattack~and our tract-level baseline on a \emph{tract}-level reconstruction with the BLOCK attribute \emph{included}. \rapattack~is initialized to the baseline. Subplots are labeled and ordered alphabetically by the state name.
    The sharp drops in the baseline match rates are explained by an anomaly in the raw PPMF microdata discussed in
    Section~\ref{sec:PPMF}, which is that in histograms
    of the number of unique rows with frequency $f$ in the overall tract, there is often a sharp spike at $f = 5$. Note that the probability
    that all $f$ copies of such a row fall into the baseline split is $1/2^f$, and that if this happens it will result in
    a non-match. For example, in the tract for AZ, there are 389 unique rows with $f = 5$ pre-split, while the counts for $f= 4$ and $f=6$ are
    88 and 27 respectively. The combination of this spike at $f=5$ and the fact that $1/32$ is relatively large results in many tracts
    having the observed dips in match rates. 
    }
    \label{fig-appx:census_tract_all_init2}
\end{figure*}

\begin{figure*}
    \centering
    \includegraphics[width=\textwidth]{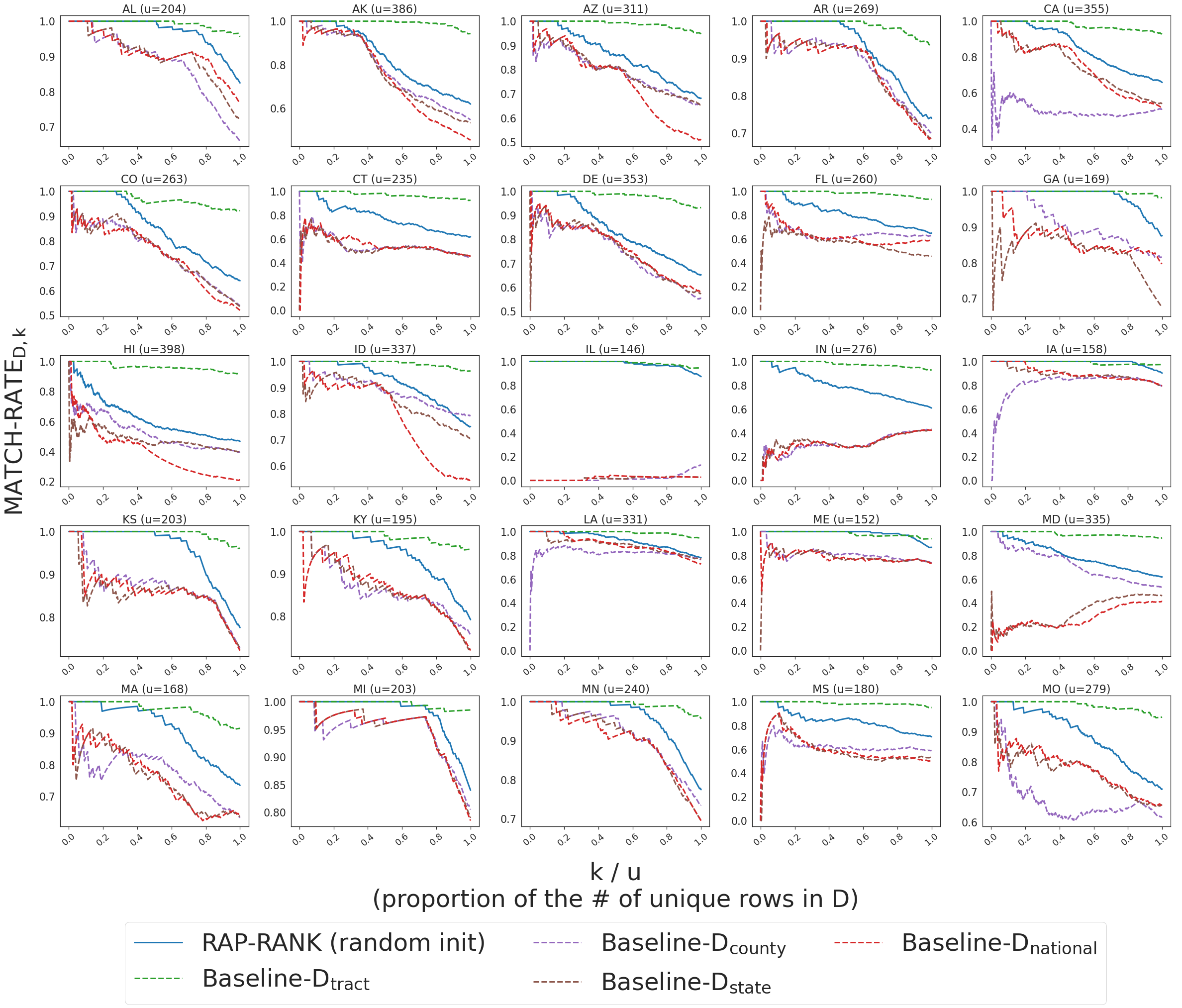}
    \caption{We plot \matchrate~of \rapattack~and our baselines on a \emph{tract}-level reconstruction with the BLOCK attribute \emph{excluded}. Subplots are labeled and ordered alphabetically by the state name.
    }
    \label{fig-appx:census_tract_ib_all1}
\end{figure*}

\begin{figure*}
    \centering
    \includegraphics[width=\textwidth]{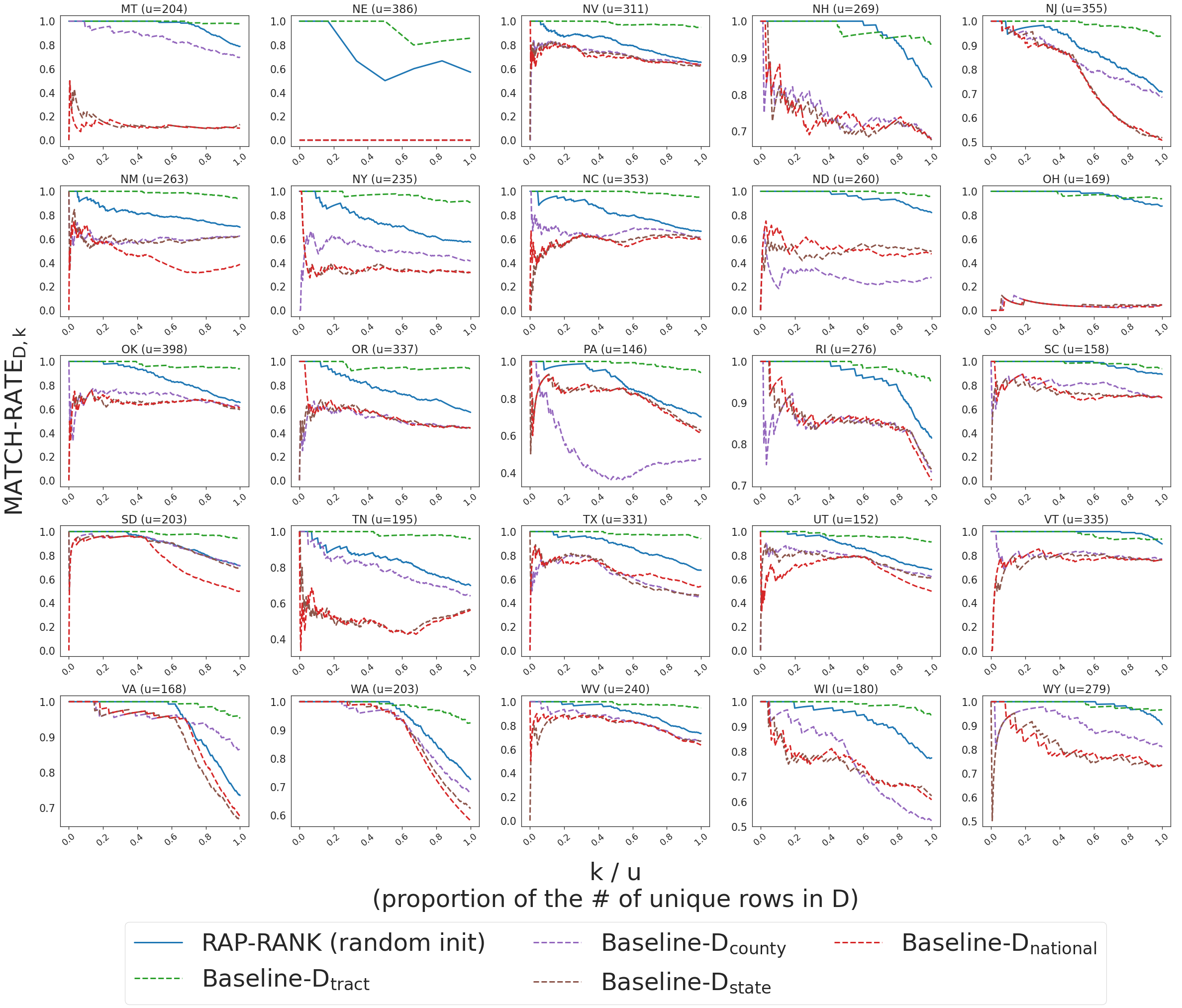}
    \caption{We plot \matchrate~of \rapattack~and our baselines on a \emph{tract}-level reconstruction with the BLOCK attribute \emph{excluded}. Subplots are labeled and ordered alphabetically by the state name.}
    \label{fig-appx:census_tract_ib_all2}
\end{figure*}

\begin{figure*}
    \centering
    \includegraphics[width=\textwidth]{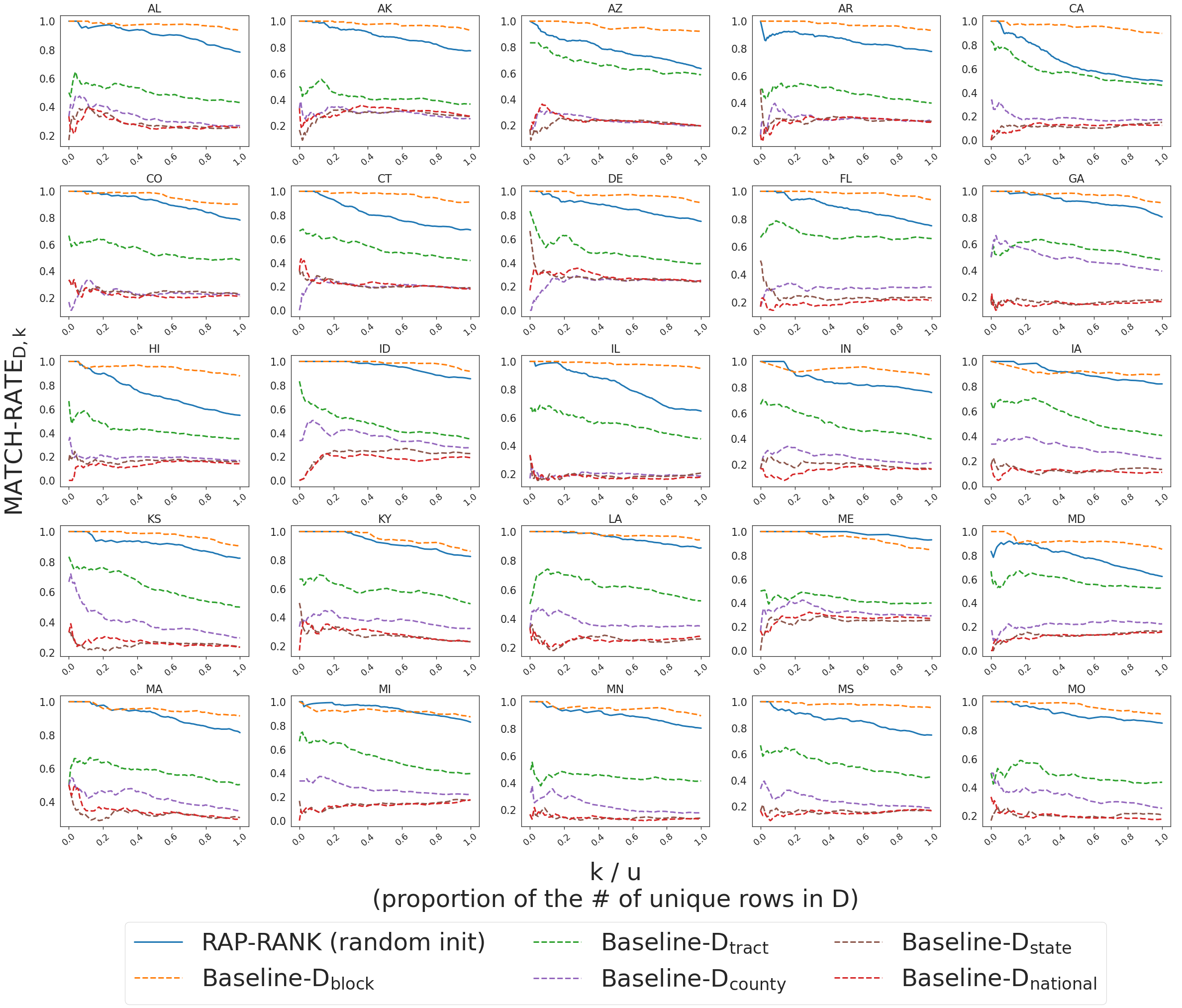}
    \caption{We plot \matchrate~of \rapattack~and our baselines on a \emph{block}-level reconstruction, where in each subplot, we average results across all blocks selected for the corresponding state. Subplots are labeled and ordered alphabetically by the state name. Since each subplot plots results average across many blocks, we do not add the value of $u$ to each subtitle.}
    \label{fig-appx:census_block_all1}
\end{figure*}

\begin{figure*}
    \centering
    \includegraphics[width=\textwidth]{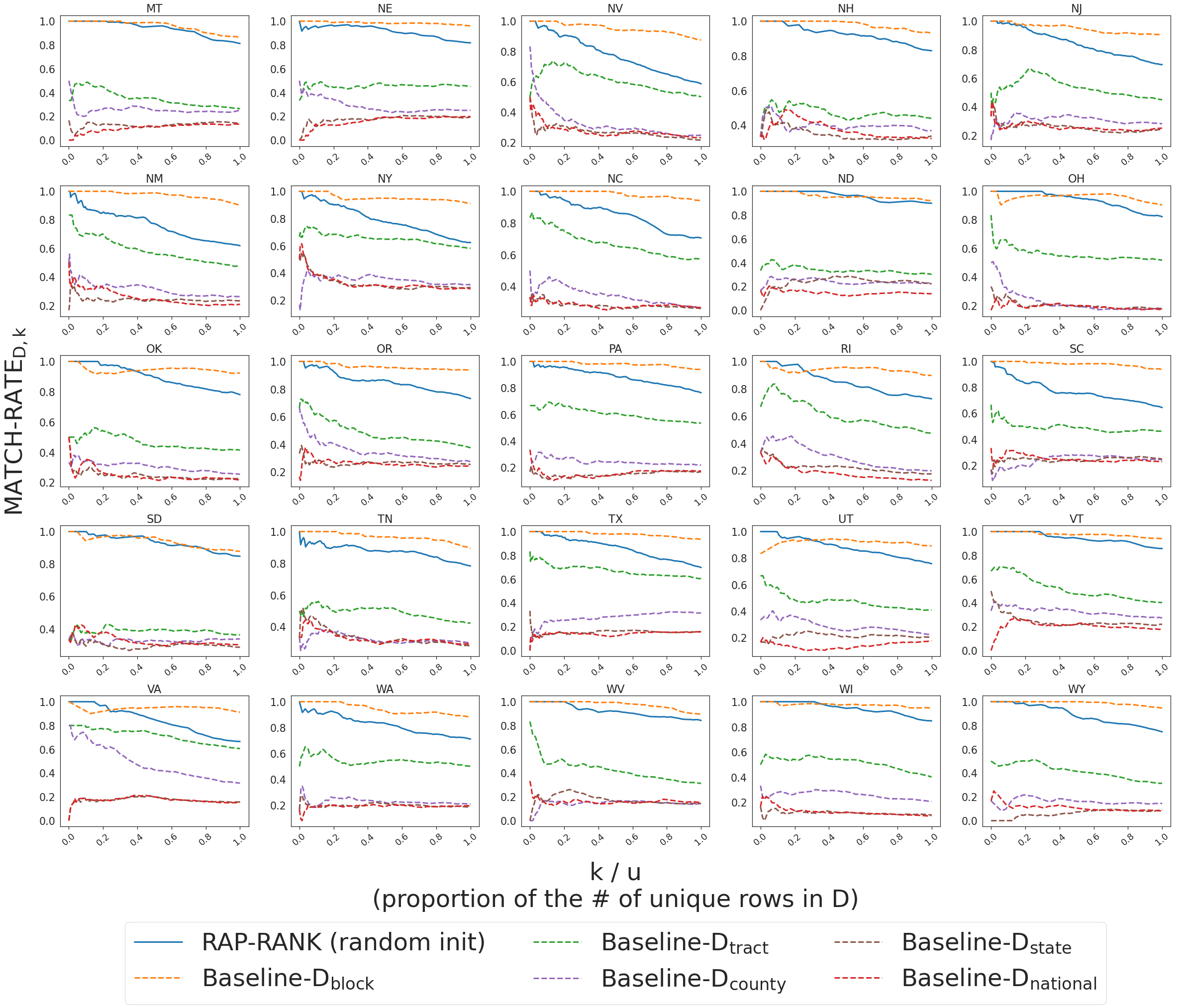}
    \caption{We plot \matchrate~of \rapattack~and our baselines on a \emph{block}-level reconstruction, where in each subplot, we average results across all blocks selected for the corresponding state. Subplots are labeled and ordered alphabetically by the state name. Since each subplot plots results average across many blocks, we do not add the value of $u$ to each subtitle.}
    \label{fig-appx:census_block_all2}
\end{figure*}

\begin{figure*}
    \centering
    \includegraphics[width=\textwidth]{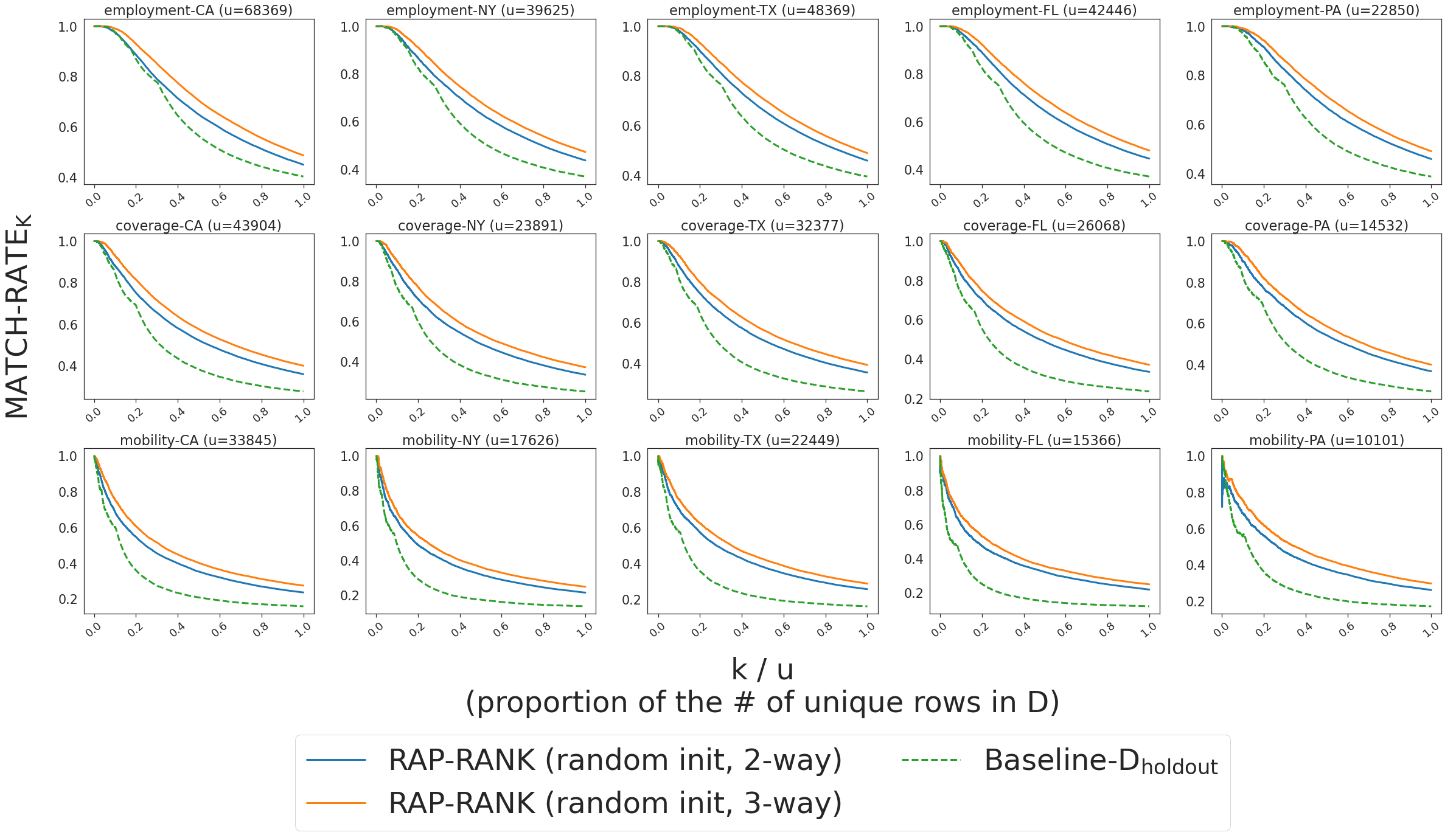}
    \caption{We plot \matchrate~of \rapattack~and our baselines on a state-level reconstruction for each state-task combination from the ACS Folktables package. We show results of \rapattack using both 2 and 3-way marginal queries.}
    \label{fig-appx:folktables_all}
\end{figure*}

In Section \ref{sec:results}, we visualized the reconstruction rates on Census and ACS datasets. However, to more easily communicate our findings, we presented results that were averaged across various geographic entities in Figures \ref{fig-main:census_avg_tract}, \ref{fig-main:census_avg_tract_init}, and \ref{fig-main:census_avg_block} for Census experiments and Figure \ref{fig-main:folktables_avg} for ACS experiments. Here, we now present more granular results. In particular, in each subplot of Figures \ref{fig-appx:census_tract_all1}, \ref{fig-appx:census_tract_all2}, \ref{fig-appx:census_tract_ib_all1}, and \ref{fig-appx:census_tract_ib_all2}, we present results for a single tract that was randomly chosen from each state, where the latter two figures (\ref{fig-appx:census_tract_ib_all1}, \ref{fig-appx:census_tract_ib_all2}) present tract-level experiments without the BLOCK feature. In addition, we plot results of \rapattack~initialized to the baseline distribution in Figures \ref{fig-appx:census_tract_all_init1} and \ref{fig-appx:census_tract_all_init2}. Like in Section \ref{sec:results}, we again average results at the block-level in Figures \ref{fig-appx:census_block_all1} and \ref{fig-appx:census_block_all2}, but now we aggregate our randomly selected blocks at the state-level. Finally, in \ref{fig-appx:folktables_all}, we present results for each of the 15 state-task combinations derived from the ACS Folktables package.

\begin{figure*}
    \centering
    \includegraphics[width=\textwidth]{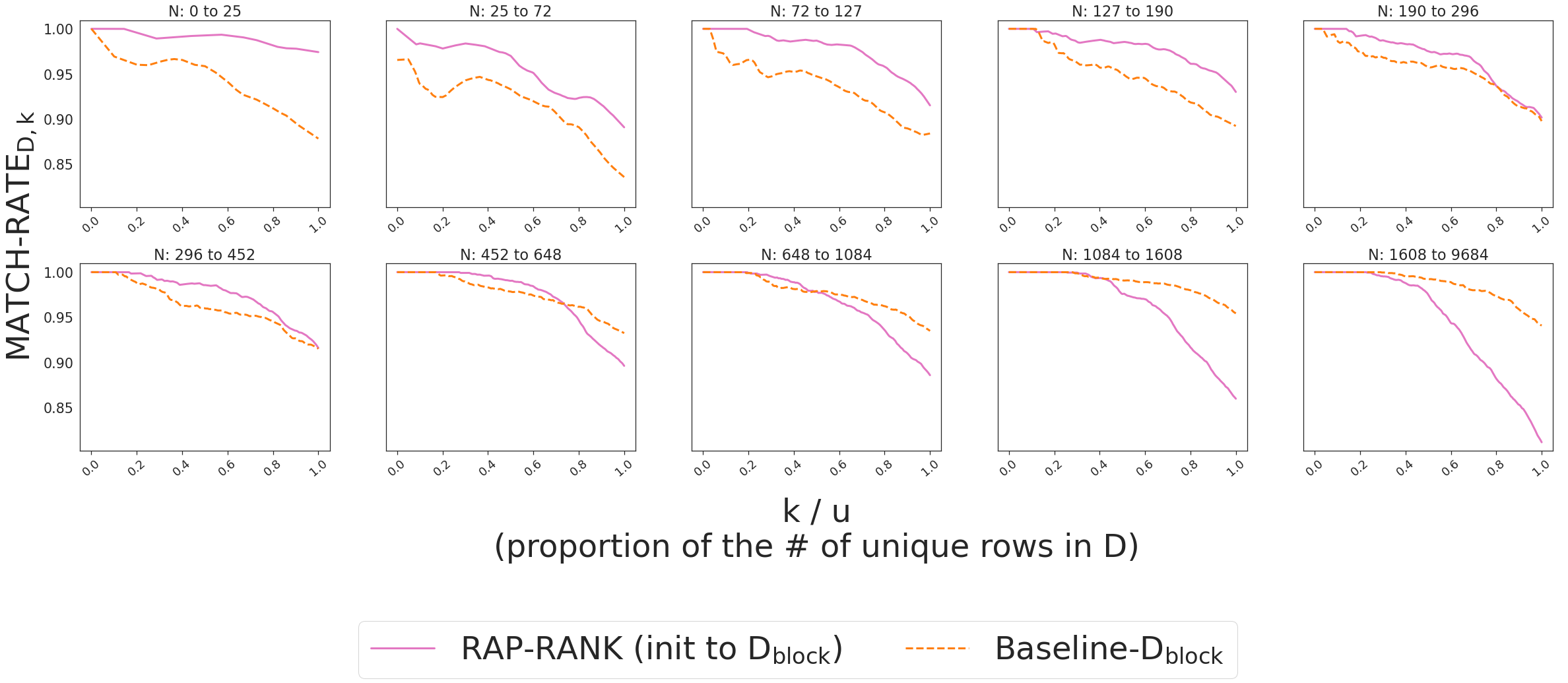}
    \caption{We plot \matchrate~of \rapattack~(initialized to the baseline) and our block-level baseline on \emph{block}-level reconstruction. We arrange the blocks into 10 subplots, ordered by size $N$ (e.g., the first subplot plots the average over the 30 smallest blocks). The title of each subplot shows the range of size $N$ for each group of blocks.}
    \label{fig-appx:ppmf_by_size}
\end{figure*}

Finally, in Figure \ref{fig-appx:ppmf_by_size}, we again plot the match rates of \rapattack~and our block-level baseline. However, to demonstrate how the performance of the two methods depends on the size of the dataset $N$, we instead group the blocks by their size. We observe that \rapattack~performs worse as $N$ increases, but conversely, the baseline performs slightly better. We attribute this phenomenon to the fact that the size of $D_{block}$ also increases with $N$, making it more likely that some element of $D_{block}$ exists in the private dataset $D$. Moreover, given that the overall data domain size of the census data is relatively small (when compared to a dataset like the ACS), the effect on the performance of our artificial baseline is immediately apparent when observing the subplots in the first row of Figure \ref{fig-appx:ppmf_by_size} where $N \le 300$

\begin{figure*}
    \centering
    \includegraphics[width=\textwidth]{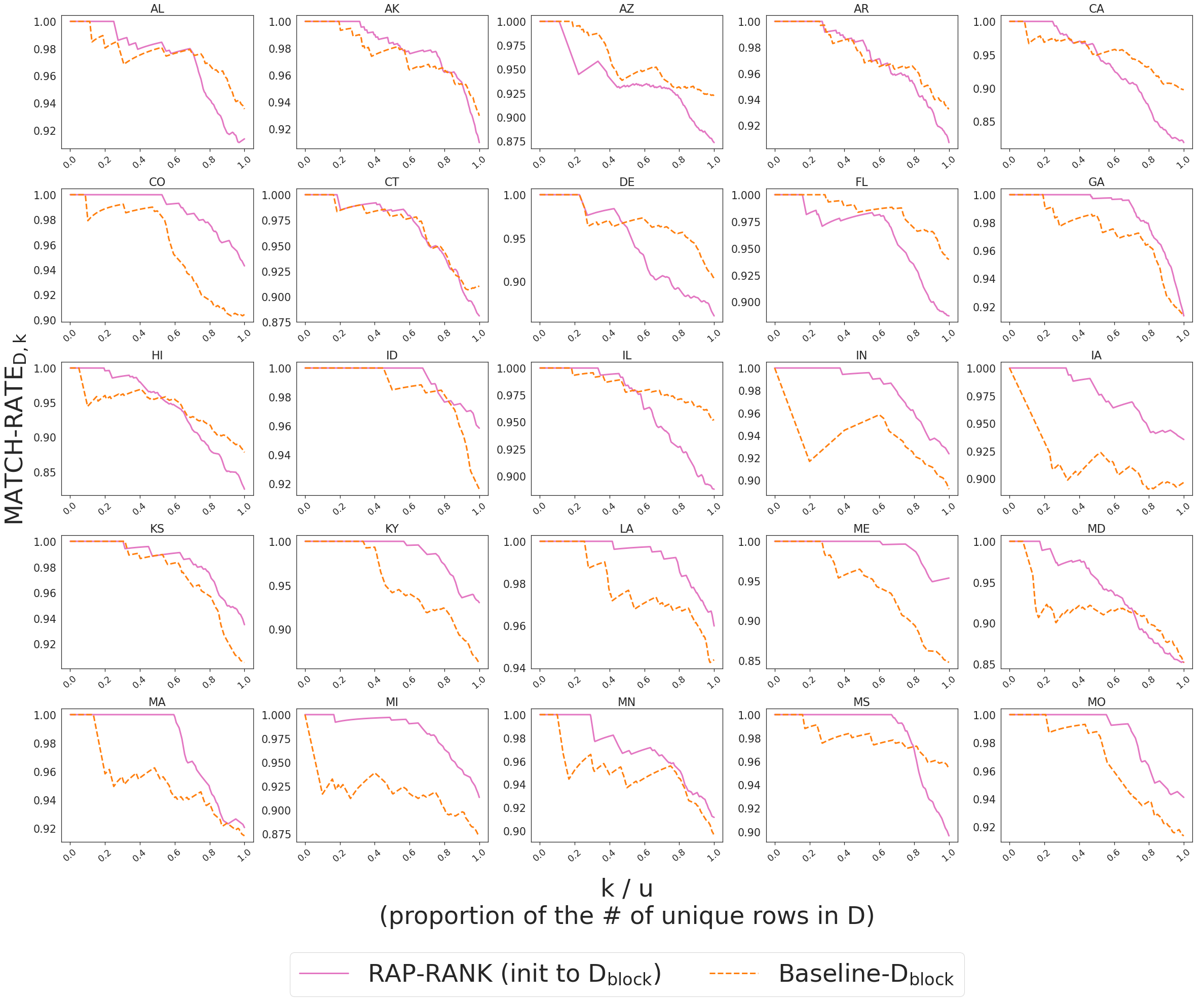}
    \caption{We plot \matchrate~of \rapattack~and our block-level baseline on \emph{block}-level reconstruction, where in each subplot, we average results across all blocks selected for the corresponding state. \rapattack~is initialized to the baseline. Subplots are labeled and ordered alphabetically by the state name. Since each subplot plots results average across many blocks, we do not add the value of $u$ to each subtitle.}
    \label{fig-appx:census_block_all_init1}
\end{figure*}

\begin{figure*}
    \centering
    \includegraphics[width=\textwidth]{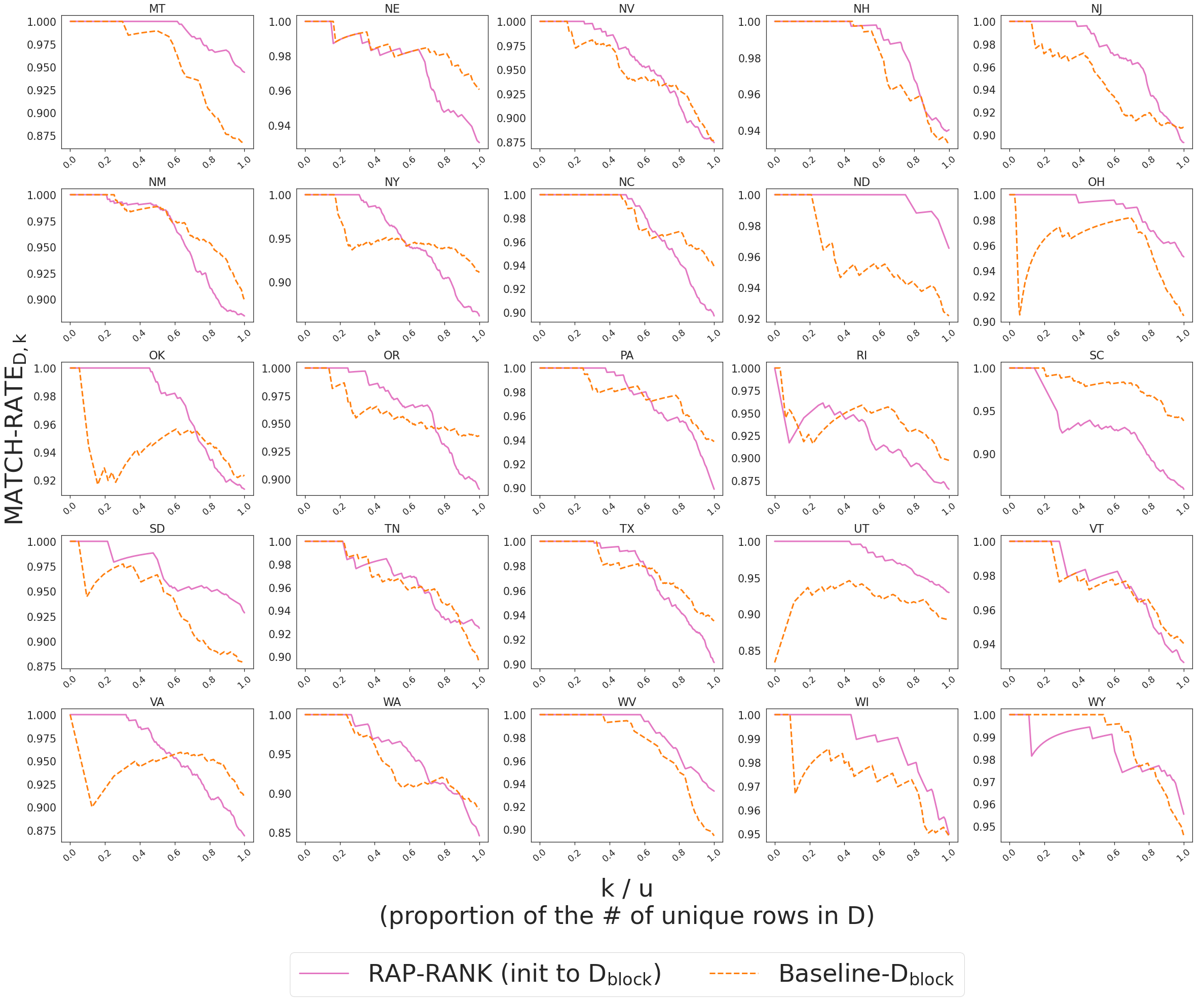}
    \caption{We plot \matchrate~of \rapattack~and our block-level baseline on \emph{block}-level reconstruction, where in each subplot, we average results across all blocks selected for the corresponding state. \rapattack~is initialized to the baseline. Subplots are labeled and ordered alphabetically by the state name. Since each subplot plots results average across many blocks, we do not add the value of $u$ to each subtitle.}
    \label{fig-appx:census_block_all_init2}
\end{figure*}

\end{document}